\documentclass[%
 aip,
 amsmath,amssymb,
 reprint,%
]{revtex4-1}

\usepackage{graphicx}
\usepackage{dcolumn}
\usepackage{bm}
\usepackage{hyperref}
\usepackage[utf8]{inputenc}
\usepackage[T1]{fontenc}
\usepackage{mathptmx}
\usepackage{color}
\usepackage{mhchem}

\usepackage{tikz}
\usepackage{amsmath}
\usepackage{etoolbox}

\newrobustcmd*{\myVtriangle}[2]{\tikz{\filldraw[draw=#1,fill=#2] (0cm,0.2cm) --
(0.2cm,0.2cm) -- (0.1cm,0cm) -- (0cm,0.2cm);}}

\newrobustcmd*{\mythickVtriangle}[2]{\tikz{\filldraw[line width=0.3mm,draw=#1,fill=#2] (0cm,0.2cm) --
(0.2cm,0.2cm) -- (0.1cm,0cm) -- (0cm,0.2cm);}}

\newrobustcmd*{\mythickErrorVtriangle}[2]{\tikz{\filldraw[line width=0.3mm,draw=#1,fill=#2] (-0.05cm,0.05cm) --
(0.05cm,0.05cm) -- (0cm,-0.05cm) -- (-0.05cm,0.05cm);  \draw[draw=#1] (0.0cm, -0.12cm) -- (0.0cm, 0.12cm) ; \draw[draw=#1] (-0.06cm, 0.12cm) -- (0.06cm, 0.12cm); \draw[draw=#1] (-0.06cm, -0.12cm) -- (0.06cm, -0.12cm)    }}

\newrobustcmd*{\mytriangle}[2]{\tikz{\filldraw[draw=#1,fill=#2] (0.0cm,0.0cm) --
(0.2cm,0cm) -- (0.1cm,0.2cm) -- (0cm,0cm);}}

\newrobustcmd*{\mysquare}[2]{\tikz{\draw[draw=#1,fill=#2] (0cm,0cm)
rectangle (0.2cm,0.2cm)}}

\newrobustcmd*{\mythicktriangle}[2]{\tikz{\filldraw[line width=0.3mm,draw=#1,fill=#2] (0.0cm,0cm) --
(0.2cm,0cm) -- (0.1cm,0.2cm) -- (0.0cm,0cm);}}

\newrobustcmd*{\mythicksquare}[2]{\tikz{\draw[line width=0.3mm,draw=#1,fill=#2] (0cm,0cm)
rectangle (0.2cm,0.2cm)}}

\newrobustcmd*{\mybarredtriangle}[2]{\tikz{\draw[draw=#1,fill=#2] (0,0) --
(0.2cm,0) -- (0.1cm,0.2cm) -- (0cm,0cm); \draw[draw=#1] (-0.1cm, 0.07cm) -- (0.3cm, 0.07cm)}}

\newrobustcmd*{\mythickbarredtriangle}[2]{\tikz{\draw[line width=0.3mm,draw=#1,fill=#2] (0,0) --
(0.2cm,0) -- (0.1cm,0.2cm) -- (0cm,0cm); \draw[draw=#1] (-0.1cm, 0.07cm) -- (0.3cm, 0.07cm)}}

\newrobustcmd*{\mybarredsquare}[2]{\tikz{\draw[draw=#1,fill=#2] (0,0)
rectangle (0.2cm,0.2cm); \draw[draw=#1] (-0.1cm, 0.1cm) -- (0.3cm, 0.1cm)}}

\newrobustcmd*{\mythickbarredsquare}[2]{\tikz{\draw[line width=0.3mm,draw=#1,fill=#2] (0,0)
rectangle (0.2cm,0.2cm); \draw[draw=#1] (-0.1cm, 0.1cm) -- (0.3cm, 0.1cm)}}

\newrobustcmd*{\mybarredcircle}[2]{\tikz{\draw[draw=#1,fill=#2] (0,0)
circle (0.1cm); \draw[draw=#1] (-0.2cm, 0.0cm) -- (0.2cm, 0.0cm)}}

\newrobustcmd*{\mythickbarredcircle}[2]{\tikz{\draw[line width=0.3mm,draw=#1,fill=#2] (0,0)
circle (0.1cm); \draw[draw=#1] (-0.2cm, 0.0cm) -- (0.2cm, 0.0cm)}}

\newrobustcmd*{\mythickErrorcircle}[2]{\tikz{\draw[line width=0.3mm,draw=#1,fill=#2] (0,0)
circle (0.06cm); \draw[draw=#1] (0.0cm, -0.12cm) -- (0.0cm, 0.12cm) ;   \draw[draw=#1] (-0.06cm, 0.12cm) -- (0.06cm, 0.12cm); \draw[draw=#1] (-0.06cm, -0.12cm) -- (0.06cm, -0.12cm)    }}

\newrobustcmd*{\mydashedline}[1]{\tikz{\draw[draw=#1] (-0.2cm, 0.2cm) -- (-0.1cm, 0.2cm); \draw[draw=#1] (-0.0cm, 0.2cm) -- (0.1cm, 0.2cm)}}

\newrobustcmd*{\mythickcross}[1]{\tikz{\draw[line width=0.3mm,draw=#1] (0,0) --
(0.2cm,0); \draw[line width=0.3mm,draw=#1] (0.1cm,-0.1cm) -- (0.1cm,0.1cm);}}

\newrobustcmd*{\mythicksidecross}[1]{\tikz{\draw[line width=0.3mm,draw=#1] (0.02928932188134524cm,-0.07071067811865477cm) --
(0.17071067811865477cm,0.07071067811865477cm); \draw[line width=0.3mm,draw=#1] (0.17071067811865477cm,-0.07071067811865477cm) -- (0.02928932188134524cm,0.07071067811865477cm);}}

\newrobustcmd*{\mybarredcross}[1]{\tikz{\draw[line width=0.3mm,draw=#1] (0,0) --
(0.2cm,0); \draw[line width=0.3mm,draw=#1] (0.1cm,-0.1cm) -- (0.1cm,0.1cm); \draw[draw=#1] (-0.1cm,0) -- (0.3cm,0);}}

\newrobustcmd*{\myline}[1]{\tikz{\draw[draw=#1] (-0.15cm, 0.1cm) -- (0.15cm, 0.1cm);\draw[line width=0.3mm,draw=#1] (-0.0cm, 0.0cm);}}

\newrobustcmd*{\mythickline}[1]{\tikz{\draw[line width=0.3mm,draw=#1] (-0.15cm, 0.1cm) -- (0.15cm, 0.1cm);\draw[line width=0.3mm,draw=#1] (-0.0cm, 0.0cm);}}

\newrobustcmd*{\mythickdashedline}[1]{\tikz{\draw[line width=0.3mm,draw=#1] (-0.2, 0.1cm) -- (-0.1cm, 0.1cm); \draw[line width=0.3mm,draw=#1] (-0.0cm, 0.1cm) -- (0.1cm, 0.1cm); \draw[line width=0.3mm,draw=#1] (-0.0cm, 0.0cm);}}

\newrobustcmd*{\mythickdasheddottedline}[1]{\tikz{\draw[line width=0.3mm,draw=#1] (-0.22, 0.1cm) -- (-0.13cm, 0.1cm); \draw[line width=0.3mm,draw=#1] (-0.085cm, 0.1cm) -- (-0.055cm, 0.1cm); \draw[line width=0.3mm,draw=#1] (-0.01cm, 0.1cm) -- (0.08cm, 0.1cm); \draw[line width=0.3mm,draw=#1] (-0.0cm, 0.0cm);}}

\newrobustcmd*{\mycircle}[2]{\tikz{\draw[draw=#1,fill=#2] (0,0)
circle (0.1cm);}}

\newrobustcmd*{\mythickcircle}[2]{\tikz{\draw[line width=0.3mm,draw=#1,fill=#2] (0,0)
circle (0.1cm);}}

\newrobustcmd*{\mydot}[1]{\tikz{\draw[line width=0.3mm,draw=#1] (0,0)
circle (0.025cm);}}

\begin{document}

\preprint{AIP/123-QED}


\title{Surface chemistry models for GaAs epitaxial growth and hydride cracking using reacting flow simulations} 



\author{Malik Hassanaly}
\email{malik[dot]hassanaly[at]nrel[dot]gov}

\affiliation{Computational Science Center, National Renewable Energy Laboratory}
\author{Hariswaran Sitaraman}
\affiliation{Computational Science Center, National Renewable Energy Laboratory}
\author{Kevin L. Schulte}
\affiliation{Chemistry \& Nanoscience Department, National Renewable Energy Laboratory}
\author{Aaron J. Ptak}
\affiliation{Chemistry \& Nanoscience Department, National Renewable Energy Laboratory}
\author{John Simon}
\affiliation{Chemistry \& Nanoscience Department, National Renewable Energy Laboratory}
\author{Kevin Udwary}
\affiliation{Kyma Technologies, Inc.}
\author{Jacob H. Leach}
\affiliation{Kyma Technologies, Inc.}
\author{Heather Splawn}
\affiliation{Kyma Technologies, Inc.}


\date{\today}

\begin{abstract}
Hydride Vapor Phase Epitaxy (HVPE) is a promising technology that can aid in the cost reduction of III-V materials and devices manufacturing, particularly high-efficiency solar cells for space and terrestrial applications. However, recent demonstrations of ultra fast growth rates ($\sim$ 500 $\mu$m/h) via uncracked hydrides are not well described by present models for the growth. Therefore, it is necessary to understand the kinetics of the growth process and its coupling with transport phenomena, so as to enable fast and uniform epitaxial growth. In this work, we derive a kinetic model using experimental data and integrate it into a computational fluid dynamics simulation of an HVPE growth reactor. We also modify an existing hydride cracking model that we validate against numerical simulations and experimental data. We show that the developed growth model and the improved cracking model are able to reproduce experimental growth measurements of \ce{GaAs} in an existing HVPE system.
\end{abstract}


\maketitle 

\section{Introduction}
\label{sec:intro}

\textit{The following article has been accepted by Journal of Applied Physics. After it is published, it will be found at \href{https://aip.scitation.org/doi/full/10.1063/5.0061222}{\underline{\color{blue}Journal of Applied Physics}}.} 
\vspace{1cm}

III-V solar cells have been used for decades in space-based applications~\cite{king2006advanced,hoheisel2010long} due to their high specific power and radiation resistance. There is an increasing push to lower the cost of these light, flexible, high-efficiency devices so that they can be used in terrestrial applications, including in transportation, building integration, and consumer electronics. Entry into these markets requires driving down the cost of III-V material manufacturing, and the recent re-emergence of hydride vapor phase epitaxy (HVPE) is one possible pathway to do so
\cite{simon2019iii}. HVPE uses relatively low-cost precursors, uses them efficiently, and recently showed growth rates for \ce{GaAs} and GaInP over 500 and 200~$\mu m/h$~\cite{mcclure2020gaas,metaferia2019gallium} respectively, which could have a significant effect on overall costs \cite{horowitz2018techno}.
It is the use of uncracked hydride precursors, in this case, \ce{AsH3} and \ce{PH3}, that enables these high growth rates using HVPE \cite{schulte2018high}. The active hydrogen released from the hydride molecule at the growth surface is theorized to scavenge chemisorbed \ce{Cl} atoms that otherwise act to inhibit growth~\cite{hollan1979fast}. This hydride-enhanced mechanism significantly reduces the kinetic barrier to growth, allowing for rates exceeding 500 monolayers/s at moderate growth temperatures ($\sim 900$~K) that are conducive to the formation of other alloys, e.g. \ce{GaInP}, needed for standard solar cell device structures. Previous work by ~\citet{denbaars1986homogeneous} showed that \ce{AsH3} could spontaneously crack at quartz surfaces following the global surface reaction 

\begin{equation}
    \label{eq:cracking}
    \ce{AsH3 -> \frac{1}{4} As4 + \frac{3}{2} H2},
\end{equation}

which they modeled with a one-step chemical mechanism.
A method to inhibit arsine cracking at the substrate is to increase the velocity of the carrier gas, thereby limiting the residence time of arsine in the hot portion of the reactor. However, larger velocities can adversely affect external gas sources mixing and lead to a non-uniform growth at the substrate, requiring careful design of the growth system. Furthermore, since surface growth is faster, the coupling with gas transport is stronger in hydride-enhanced HVPE growth. In this context, a simulation-based understanding that couples the chemical kinetics of arsine cracking and \ce{GaAs} deposition with transport phenomena is paramount for the optimization of the overall manufacturing process.

Existing growth models for the deposition of \ce{GaAs} in the literature only consider deposition from arsenic vapor (\ce{As2} or \ce{As4}) \cite{shaw1975kinetic,kangawa2002theoretical}, but not from uncracked \ce{AsH3}. In this work, a one-step kinetic model for the deposition of \ce{GaAs} from uncracked arsine and gallium chloride is developed and describes the following global reaction

\begin{equation}
    \ce{AsH3 (g) + GaCl (g) \rightleftharpoons GaAs (s) + HCl (g) + H2 (g)}. 
\end{equation}

The main contributions of the paper are: a) the development of a kinetic model for surface growth of uncracked arsine; b) uncertainty bounds on the growth model parameters are provided using a Bayesian calibration approach; c) an update of the arsine cracking model of \citet{denbaars1986homogeneous}; d) the cracking and growth models are integrated into a computational fluid dynamics (CFD) solver.

The surface growth model is developed and validated by modeling the reaction kinetics and the fluid dynamics in a lab-scale reactor designed for a 2-inch substrate, and for which experimental growth data is available. The rest of the paper is organized as follows. Section~\ref{sec:config} presents the lab-scale reactor simulated and the dataset used to develop the kinetic model. The numerical method is described in Sec.~\ref{sec:numericalmethod} and the calibration of the cracking model from experimental data is shown in Sec.~\ref{sec:cracking}. In Sec.~\ref{sec:model}, the finite rate kinetic model for \ce{GaAs} deposition is developed. Concluding remarks are provided in Sec.~\ref{sec:conclusions}.

\section{Configuration and experimental dataset}
\label{sec:config}

\subsection{D-HVPE reactor}

\begin{figure}[ht!]
\centering
\includegraphics[width=0.47\textwidth,trim={0cm 1cm 0cm 1cm},clip]{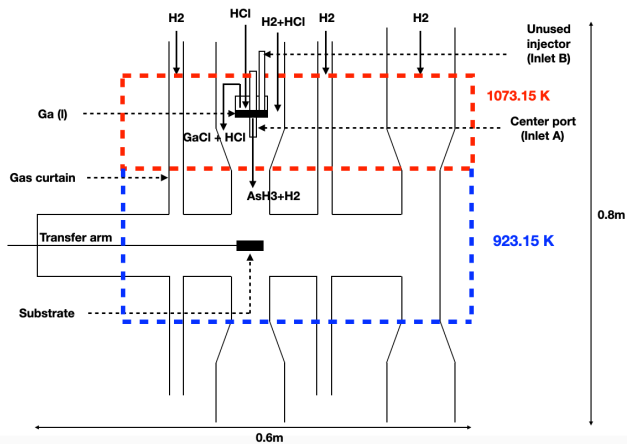}
\caption{Schematic of the D-HVPE reactor. Solid arrows indicate flows, while dashed arrows point to different elements of the reactor. An external furnace maintains part of the reactor wall temperature at $1073.15$~K (area surrounded by \mythickdashedline{red}) and the rest at $923.15$~K (area surrounded by \mythickdashedline{blue}).}
\label{fig:illustrationDHVPE}
\end{figure}

The configuration simulated in this work (see illustration in Fig.~\ref{fig:illustrationDHVPE}) is from experimental studies by \citet{schulte2018high} that used a Dynamic-Hydride Vapor phase epitaxy (D-HVPE) reactor where growth rates with uncracked arsine were also reported.
These experiments were conducted with a Dynamic-Hydride Vapor phase epitaxy (D-HVPE) reactor \cite{young2013high} which addresses manufacturing challenges posed by HVPE. Highly efficient thin-film solar cells typically consist of several layers, each one made of a different III-V compound \cite{biefeld2015science}. When the growth rate of the solar cell is low enough and the reactor equilibrates quickly, the substrate can be kept idle in the same chamber while different external gas are injected over time. With a higher growth rate and large residence time of gases - such as in HVPE deposition reactors - this strategy can lead to non-abrupt interfaces between the layers, which is detrimental to solar cell efficiency. In the D-HVPE, the aforementioned issue is avoided by physically moving the substrate to a different chamber each time a new layer needs to be grown. In order to prevent inter-chamber flow contamination, the chambers are separated by vertical gas curtains created by a flow of H2 (see illustration in Fig.~\ref{fig:illustrationDHVPE}).

The D-HVPE reactor considered here has been experimentally \cite{metaferia2019gallium,schulte2018high,schulte2019uniformity} and numerically \cite{mcclure2020gaas,schulte2019uniformity} shown to provide high growth rate and efficiency, while potentially enabling in-line manufacturing.

The metal-halide is generated in-situ by injecting \ce{HCl} into a boat/reservoir of liquid \ce{Ga}. The boat is externally heated at $1073.15$~K to drive the kinetics of the \ce{HCl + Ga -> GaCl} reaction. The boat conversion efficiency was previously estimated to be 90\% \cite{schulte2016kinetic}. It outputs a mixture of gaseous gallium chloride \ce{GaCl} and residual \ce{HCl} that were not converted. A center port (called inlet A in \citet{schulte2018high}) introduces a mixture of \ce{AsH3} and \ce{H2} in the chamber. The reactants are then deposited on a substrate externally heated at $923.15$~K. In Fig.~\ref{fig:illustrationDHVPE}, the external heating is represented by the red and blue dashed boxes.

\subsection{Dataset}

The dataset used for validation of our model is that presented in \citet{schulte2018high} where growth rates at the center of the substrate were measured under a variety of flow conditions. The volume flow rates at standard pressure and temperature for all the cases considered are shown in Tab.~\ref{tab:conditions}. The experiments were conducted at a pressure of $0.829$~atm and with external gas injected at ambient temperature.

\begin{table}[ht!]
\setlength{\abovecaptionskip}{0pt}
\setlength{\belowcaptionskip}{5pt}
\caption{Volume flow rates in standard centimeter cubes per minute (sccm) for each case simulates. ``Top Chamber" denotes the top boundary of the chamber that contains the substrate. ``Adjacent Chamber" denotes the top boundary of the chamber that does not contain the substrate.}
\label{tab:conditions}
\begin{ruledtabular}
\begin{tabular}{|c|c|c|c|c|c|c|c|}
\hline
 [sccm] &
\multicolumn{2}{c|}{Center Port} &
\multicolumn{3}{c|}{Top Chamber} &
\multicolumn{1}{c|}{Adjacent Chamber} &
\multicolumn{1}{c|}{Curtains} \\ \hline
Case &
\ce{H2} &
\ce{AsH3} &
\ce{H2} &
\ce{HCl} &
\ce{GaCl} &
\ce{H2} &
\ce{H2} \\ \hline
Case 1 & 1000 & 44.5 & 8200 & 21 & 9 & 10000 & 2500 \\ \hline
Case 2 & 1600 & 44.5 & 7600 & 21 & 9 & 10000 & 2500 \\ \hline
Case 3 & 2200 & 44.5 & 7000 & 21 & 9 & 10000 & 2500 \\ \hline
Case 4 & 2500 & 33.5 & 7500 & 20.8 & 7.2 & 10000 & 2500 \\ \hline
Case 5 & 2500 & 33.5 & 7500 & 21 & 9 & 10000 & 2500 \\ \hline
Case 6 & 2500 & 33.5 & 7500 & 21.2 & 10.8 & 10000 & 2500 \\ \hline
Case 7 & 2500 & 33.5 & 7500 & 21.6 & 14.4 & 10000 & 2500 \\ \hline

\end{tabular}
\end{ruledtabular}
\end{table}

\begin{figure}[ht!]
\centering
\includegraphics[width=0.47\textwidth]{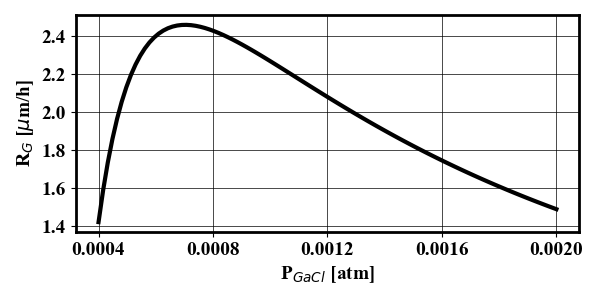}
\caption{\ce{GaAs}(s) growth rate via \ce{As4} reaction pathway at $923.15$~K for various partial pressure of \ce{GaCl}.}
\label{fig:as4pathway}
\end{figure}

An important observation from the experiment was the near-zero growth of \ce{GaAs} when \ce{AsH3} was recirculated into the reactor (through inlet B in Fig.\ \ref{fig:illustrationDHVPE}) similar to \ce{GaCl} rather than direct injection through the center port (inlet A in Fig.\ \ref{fig:illustrationDHVPE}). The larger residence time of \ce{AsH3} in this case led to increased cracking at the reactor walls before reaching the substrate. This hypothesis is further bolstered by the success of growth rate models that considered a deposition pathway only via cracked arsine \cite{schulte2019uniformity}. Under the conditions where all the arsine has cracked and in the range of operating conditions investigated, an equilibrium kinetic model \cite{schulte2016kinetic} indeed predicts growth rates of at most $2$~$\mu$m/h (Fig.~\ref{fig:as4pathway}) which is at least one order of magnitude lower than growth rates measured in \citet{schulte2018high}. Here, the \ce{As4} pathway is inhibited by the low residence time of external gas sources and the low substrate temperature. Therefore, the growth rate available in \citet{schulte2018high} is particularly suited to deduce a kinetic model for the \ce{AsH3}-pathway for epitaxial growth.

\section{Numerical method}
\label{sec:numericalmethod}

To derive a kinetic model for the surface reaction via uncracked hydrides, computational fluid dynamics (CFD) simulations of the experimental configuration are conducted. The kinetic model depends on parameters that are then calibrated to match experimental growth rates. The fluid solver used is described in App.~\ref{app:fluidSolver}, the implementation of surface reactions is explained in App.~\ref{app:numericsSurfaceReaction} and the transport and thermodynamic parameters are given in App.~\ref{app:thermoParam}. 

The boat injector is omitted from the domain to simplify the geometry. Instead, the boat injector is modeled using an injector spread over the top of the chamber aligned with the substrate, assuming uniform composition of \ce{GaCl}, \ce{HCl} and \ce{H2}. The volume flow rates for the top chamber are available in Tab.~\ref{tab:conditions}. Additionally, to reduce the range of scales resolved, the inner region of the center port is not meshed and its exit is treated as an inlet boundary condition. Because \ce{AsH3} is mostly exposed to walls inside the center port, this is where it mostly cracks. The boundary conditions at the exit of the center port need specific treatment to incorporate this effect and are described in App.~\ref{app:surrogateModelCenterPort}.

\begin{figure}[ht!]
\centering
\includegraphics[width=0.47\textwidth]{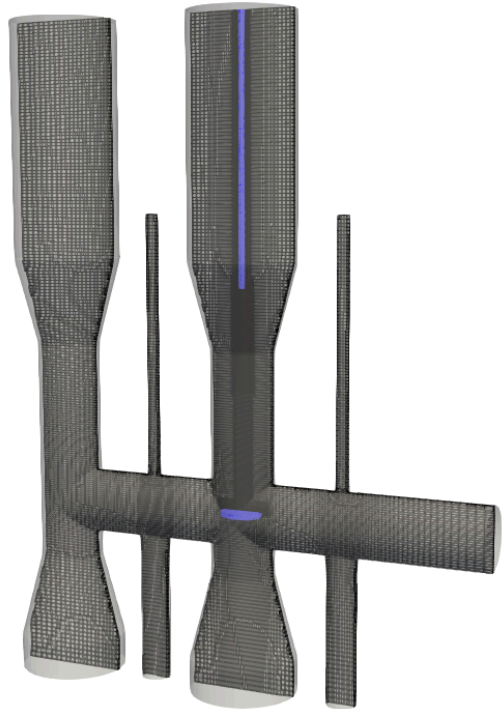}
\caption{Illustration of the mesh used. The blue parts denote the center port and the substrate. Note the refinement near the center port exit and near substrate, where gradients are the largest.}
\label{fig:mesh}
\end{figure}

A hexahedral-dominant mesh shown in Fig.~\ref{fig:mesh} is used with refinement near the injector exit and the platter, where large gradients of velocity and composition are observed. To ensure that the results presented are not influenced by numerical errors, two different meshes are used by increasing the resolution in the chamber that contains the substrate. The coarse mesh contains about $0.5$ million computational cells and the finer mesh consists of $1$ million cells. The smallest mesh size in all three directions is $0.75$~mm and $0.47$~mm for the coarse and fine meshes, respectively.

\section{Arsine cracking model}
\label{sec:cracking}

As mentioned in Sec.~\ref{sec:intro}, the cracking of hydrides (here \ce{AsH3}) critically impacts the growth rate of \ce{GaAs} layers. Therefore, a cracking kinetic model needs to be included with other surface reactions and coupled with the transport equations to adequately predict deposition rates. A global kinetic model for the decomposition of hydrides at quartz surfaces (the walls of the D-HVPE reactor considered here are also made of quartz) of \ce{AsH3} diluted within a \ce{H2} carrier gas was developed in \citet{denbaars1986homogeneous}. One of the experiments reported in \citet{denbaars1986homogeneous} included measurement of \ce{AsH3} concentration after it has passed through a $5$~cm diameter and $12.5$~cm long heated quartz tube at a speed of $1.25$~cm/s. The kinetic rate of the global reaction (Eq.~\ref{eq:cracking}) was modeled in the Arrhenius form as 

\begin{equation}
    K_{cracking} = A_{cracking} exp\left(-\frac{Ea_{cracking}}{R_u T}\right),
\end{equation}

where $A_{cracking}$ and $Ea_{cracking}$ are constant kinetic coefficients for pre-exponential factor and activation energy, $R_u$ is the universal gas constant, and $T$ is the surface temperature. The values reported for $K_{cracking}$ by \citet{denbaars1986homogeneous} were computed by assuming with respect to volumetric arsine concentration, given by

\begin{equation}
     K_{cracking} = \frac{1}{\Delta t} log \left(\frac{C_0}{C_f}\right),
\end{equation}

where $\Delta t$ is the residence time of the gas in the quartz tube, $C_0$ and $C_f$ are the volumetric molar concentrations of \ce{AsH3} at the inlet and outlet of a tube through which a mixture of arsine and hydrogen flowed. While the expression of $\Delta t$ is not explicitly given, the value of $K_{cracking}$ provided in \citet{denbaars1986homogeneous} suggests that it was calculated as $\frac{L}{U}$, where $L$ is the length of the tube and $U$ is the streamwise velocity. The expression chosen for $K_{cracking}$ neglected the effect of surface-to-volume ratio which depends on the geometry of the tube. For instance, if the tube had a smaller diameter, and under the same conditions as the experiments -- in particular same volume flow rates --, the apparent $K_{cracking}$ would be larger than the one seen in the experiments. The importance of including geometry considerations in the kinetic constants was also recognized elsewhere \cite{harrous1988phosphine}. Furthermore, since $K_{cracking}$ is the rate constant for a surface reaction it should have units of m.s$^{-1}$. \citet{denbaars1986homogeneous} factored in the area-to-volume ratio (units of $m^{-1}$) within $K_{cracking}$ that gave rise to a first-order rate constant with units of $1/s$. The authors reported a value for $Ea_{cracking}$ = 34$~kcal/mol$ by varying the temperature of the tube but did not propose a value for $A_{cracking}$.
One of the main contributions of this work is the calibration of $A_{cracking}$ along with a correction to the expression for $K_{cracking}$ that accounts for the surface-to-volume ratio.

One can estimate the cracking of \ce{AsH3} in a cylindrical tube using a steady one dimensional advection-reaction equation (justification is provided in App.~\ref{app:surrogateModelCenterPort}). The mass fraction of arsine $Y_{\ce{AsH3}}$ along the streamwise direction follows the relation

\begin{equation}
    \pi R^2 \frac{d}{dx} \left(\rho U Y_{\ce{AsH3}}\right) = -K_{cracking} \rho Y_{\ce{AsH3}} 2 \pi R,
\end{equation}

where $U$ is the streamwise velocity and $R$ is the tube radius. Assuming that the density  $\rho$ varies only due to temperature variations, one can derive an approximate relation for the ratio $C_f/C_0$ as 

\begin{equation}
    K_{cracking}  = \frac{U R}{ 2 L }log\left(\frac{C_0}{C_f}\right).
\end{equation}

The new pre-log factor can be thought of as a residence timescale rescaled with a non-dimensional ratio of catalytic surface per unit volume. With that rescaling, the cracking kinetic coefficient now includes the effect of tube radius and the expression can be directly used to obtain a value for $A_{cracking}$ by fitting the experimental data to the 1D model. 

\begin{figure}[ht!]
\centering
\includegraphics[width=0.45\textwidth]{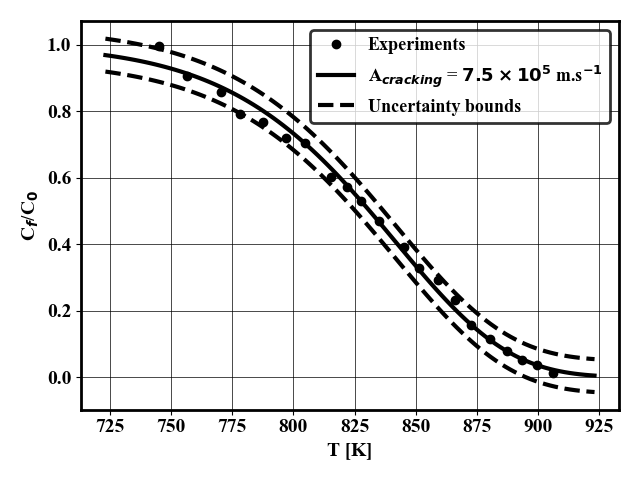}
\includegraphics[width=0.45\textwidth,trim={3.5cm 7cm 3.5cm 7cm},clip]{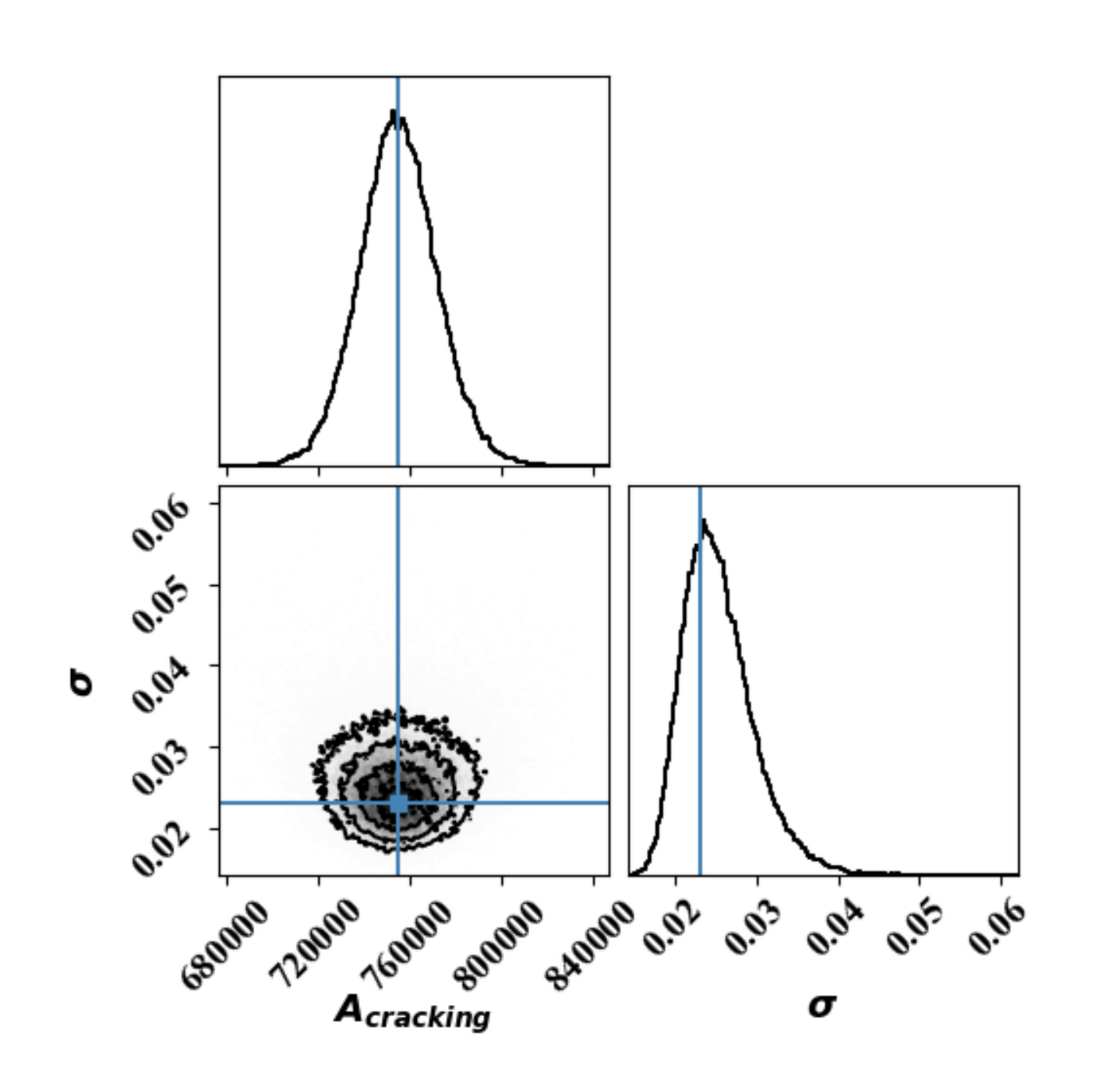}
\caption{Top: experimental measurement (\mythickcircle{black}{black}) of \ce{AsH3} cracking at different temperatures, 1-step cracking model prediction with $A_{cracking}=7.5 \times 10^5$~m.s$^{-1}$ (\mythickline{black}) corresponding to the largest posterior probability and uncertainty bounds $\pm \sigma$ (\mythickdashedline{black}). Bottom: corner plot of the 2-dimensional posterior PDF.}
\label{fig:fitdenbaars}
\end{figure}

To incorporate modeling and experimental uncertainty in the calibration of $A_{cracking}$, a Bayesian calibration approach is adopted~\cite{braman2013bayesian,bell2019bayesian}. The output of that procedure is a posterior probability density function (PDF) of $A_{cracking}$ given the experimental data available and is computed as

\begin{equation}
    P(A_{cracking}|D) \sim P(D|A_{cracking}) P(A_{cracking}),
\end{equation}

where $P(A_{cracking}|D)$ is the posterior PDF, $P(A_{cracking})$ is a prior PDF (describes the knowledge of $A_{cracking}$ without any experimental measurement) and $P(D|A_{cracking})$ is the likelihood of the data. An uninformative prior $P(A_{cracking}) \sim \mathcal{U}(10^4,2\times10^7)$ is chosen. Following \citet{braman2013bayesian}, a Gaussian likelihood is adopted,

\begin{equation}
    P(D|A_{cracking}) \sim \frac{1}{(2 \pi \sigma)^{N_d/2}} exp\left[ -\frac{1}{2 \sigma^2}  \sum_{i=1}^{N_d} (x_i(A_{cracking}) - D_i)^2 \right],
\end{equation}

where $N_d$ is the number of data points, $x_i$ is the model prediction for the $i^{th}$ experiment, $D_i$ is the $i^{th}$ experimental value, and $\sigma$ is a hyperparameter which describes the experimental errors. Since $\sigma$ is unknown here, it is fit as part of the Bayesian calibration procedure with prior $\mathcal{U}(0,10)$, along with $A_{cracking}$ \cite{braman2013bayesian}. It was found that the choice of prior had little influence on the calibrated parameters (shown in App.~\ref{app:bayesianPrior}. The results of the Bayesian calibration are shown in Fig.~\ref{fig:fitdenbaars}.

\section{Gallium arsenide growth model}

\label{sec:model}

\subsection{Model functional form}

The surface reaction is modeled as an irreversible forward reaction 

\begin{equation}
    \ce{AsH3 (g) + GaCl (g) -> GaAs (s) + HCl (g) + H2 (g)}. 
\end{equation}

The assumption of irreversibility is justified by its large equilibrium constant which suggests that the backward reaction is unlikely \cite{mcclure2020gaas,gruter1989deposition}. A finite rate chemistry model is developed hereafter for the forward reaction. An infinitely fast model is also constructed and the results are reported in App.~\ref{app:infiniteChemistry}.

The finite rate chemistry model for the reaction at the substrate is constructed by assuming the Arrhenius form $K_{reac} = A_{reac} T^{\beta_{reac}} exp(-Ea_{reac}/R_u T)$, where $A_{reac}$, $\beta_{reac}$ are kinetic constants and $Ea_{reac}$ is the activation energy. In the experiments conducted in \citet{schulte2018high}, it was found that for the temperature range $[873-953~K]$, the dependence of the epitaxial growth rate with temperature was negligible. Therefore, $K_{reac}$ is assumed to be independent of temperature over similar temperature ranges studied in this work.

The objective of this section is to find a value of $K_{reac}$. The CFD model is used to simulate the D-HVPE reactor with different values of $K_{reac}$. The resulting growth rates are obtained by computing the molar flux of \ce{AsH3} at the substrate, which is equal to the number of moles of \ce{GaAs} produced. The density of \ce{GaAs}(s) is then used to compute the growth rate obtained via CFD, which is then compared to the ones reported in the \citet{schulte2018high}. The growth rates are measured at the center of the substrate.

\subsection{Model calibration}
\label{sec:calibrationMethod}

\begin{figure*}[ht!]
\centering
\includegraphics[width=0.80\textwidth,trim={0cm 5cm 0cm 5cm},clip]{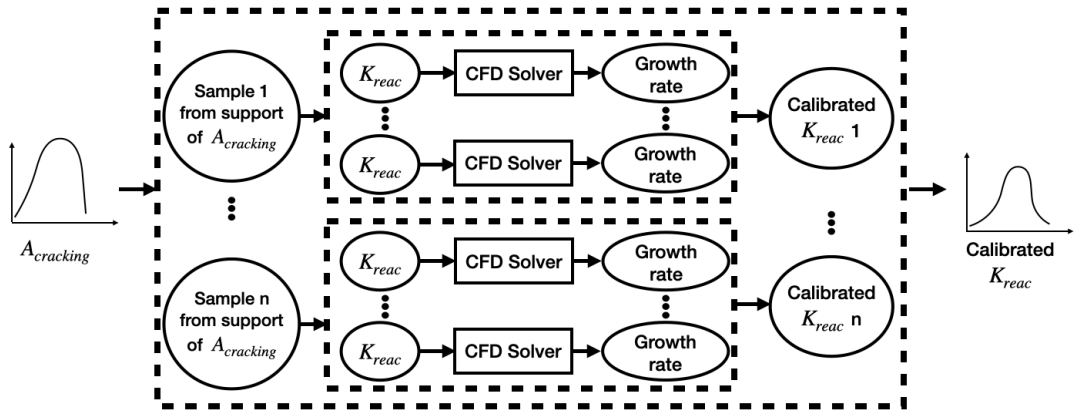}
\caption{Illustration of the workflow of the calibration procedure.}
\label{fig:calibrationProcedure}
\end{figure*}

Eleven values of $K_{reac}$ are chosen within the range [$45$~m.s$^{-1}$,$310$~m.s$^{-1}$]. For each value of $K_{reac}$, the seven cases shown in Tab.~\ref{tab:conditions} are simulated. In absence of uncertainty, a total of $77$ numerical simulations would be needed for calibration. In the present case, the optimal value of $K_{reac}$ depends on the value of $A_{cracking}$ obtained by fitting the cracking model. To account for uncertainty in the value of $A_{cracking}$, the calibration procedure is conducted for ten values of $A_{cracking}$ chosen in the range [$6.5\times10^5$~m.s$^{-1}$,$8.5\times10^5$~m.s$^{-1}$], which span the support of its posterior distribution. The function that related $K_{reac}$ and $A_{cracking}$ is then convolved with the posterior distribution of $A_{cracking}$ to deduce the PDF of $K_{reac}$. The overall procedure is illustrated in Fig.~\ref{fig:calibrationProcedure} and requires a total of $770$ runs.

Due to the large number of simulations required, the calibration is done with the coarse grid ($0.5$ million computational cells). The fine mesh ($1$ million computational cells) is used afterwards to verify grid convergence for the estimated values of $K_{reac}$.

\subsection{Results}

\begin{figure}[ht!]
\centering
\includegraphics[width=0.35\textwidth,trim={0cm 3cm 0cm 3cm},clip]{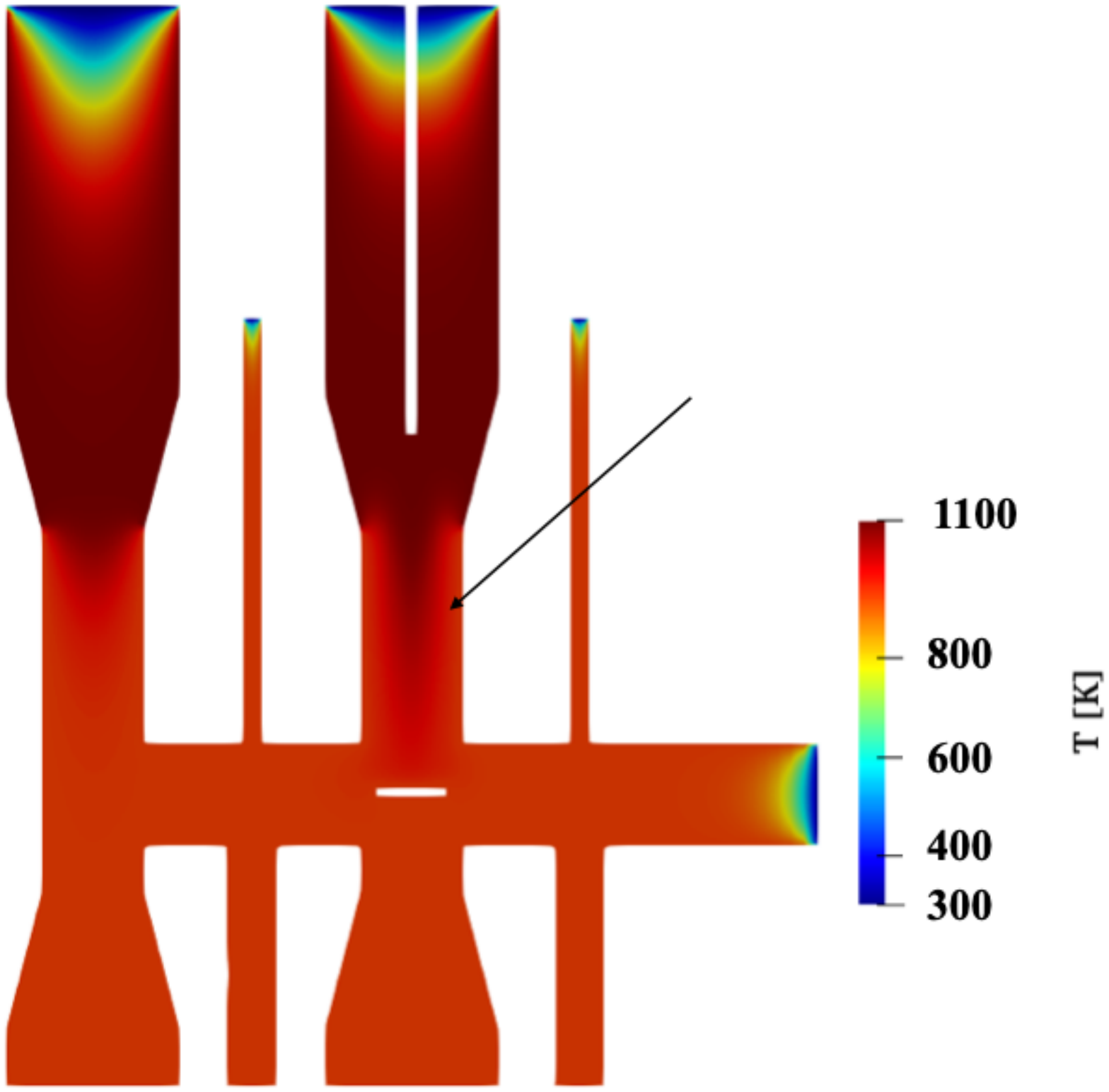}
\includegraphics[width=0.35\textwidth,trim={3.1cm 0cm 3cm 0cm},clip]{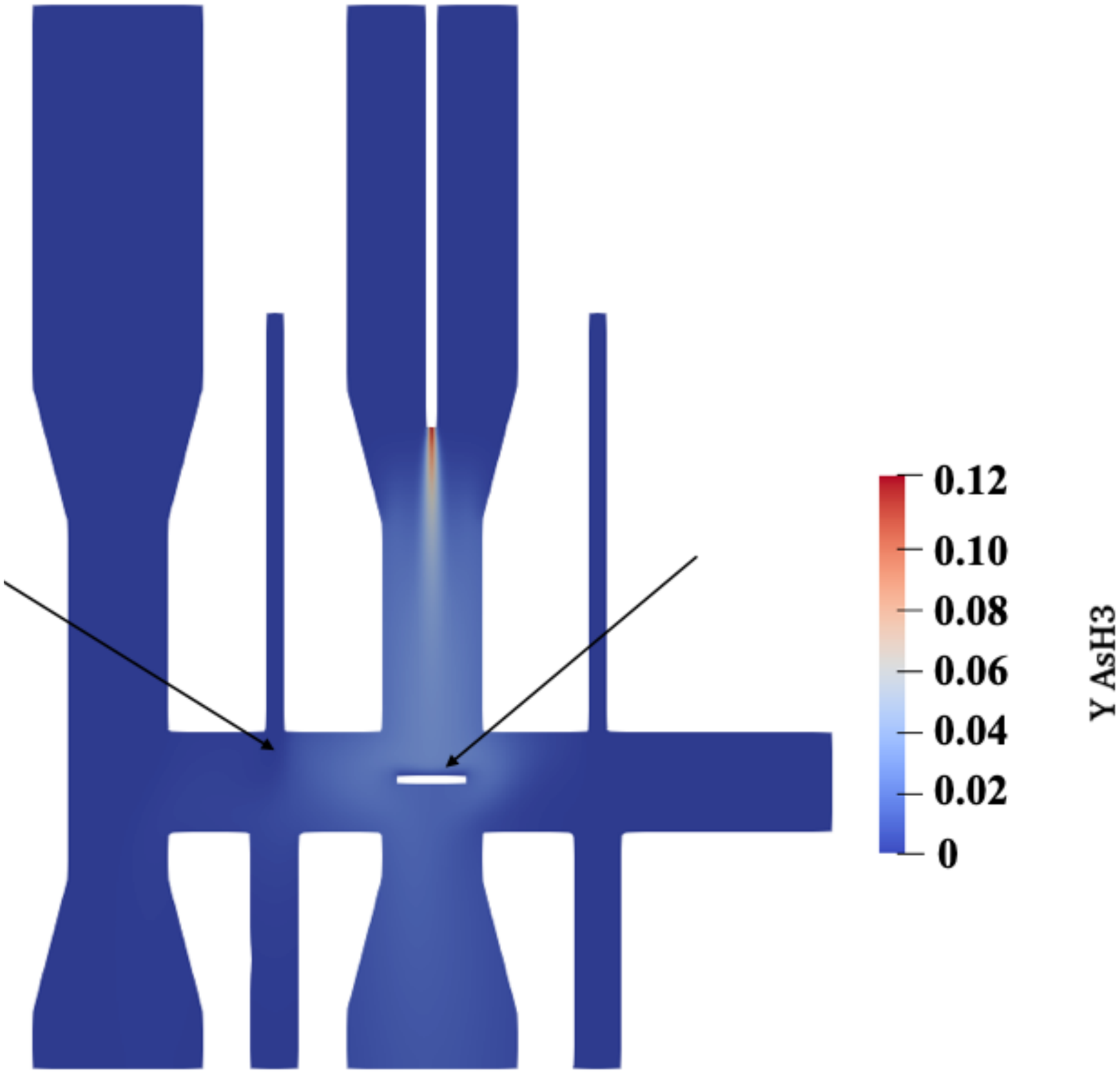}
\includegraphics[width=0.35\textwidth,trim={3.1cm 0cm 3cm 0cm},clip]{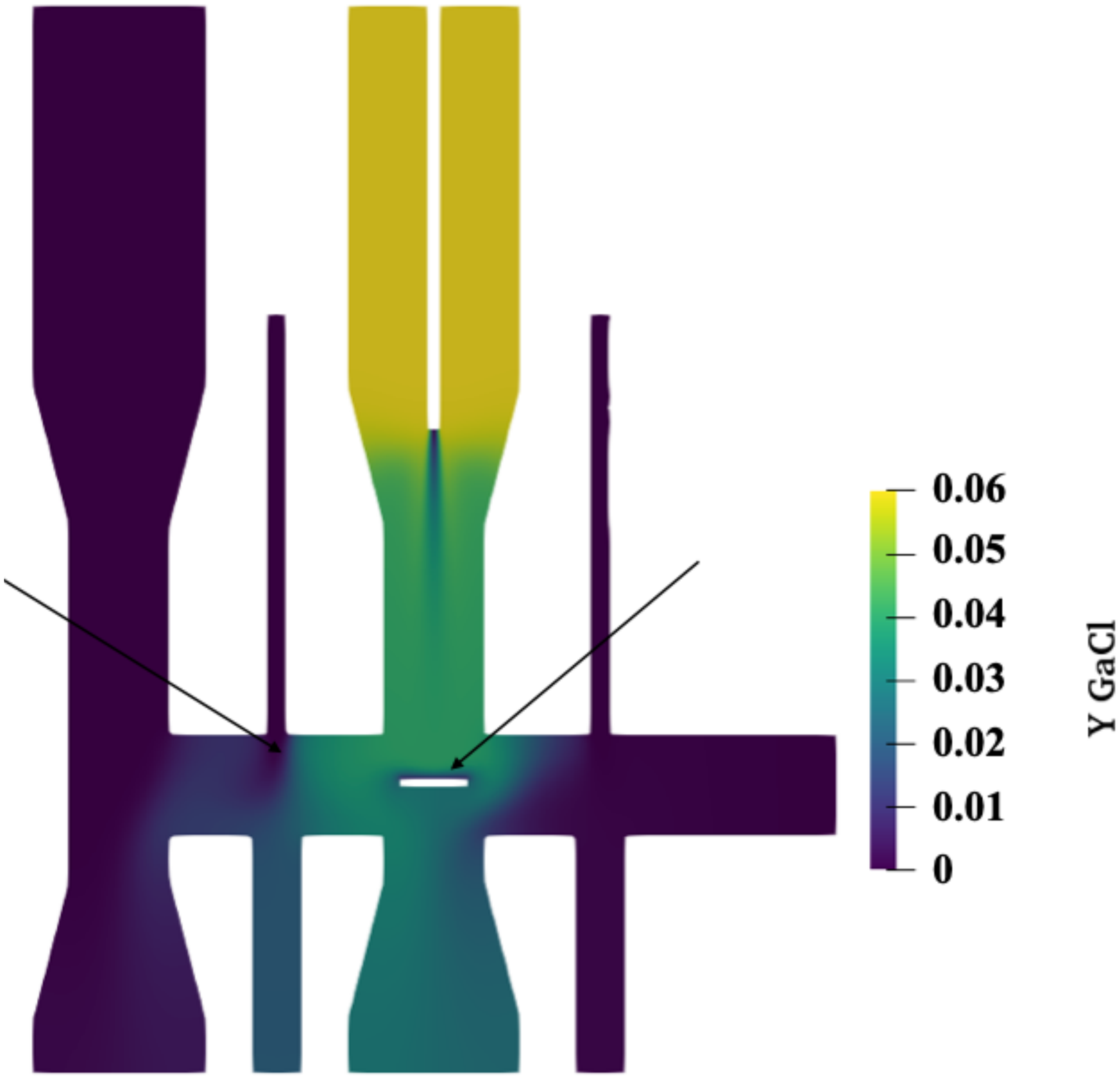}
\caption{Top: steady-state contour of temperature in the D-HVPE reactor for Case 3. Middle: steady-state contour of mass fraction of \ce{AsH3} in the D-HVPE reactor for Case 3. Bottom: steady-state contour of mass fraction of \ce{GaCl} in the D-HVPE reactor for Case 3. Arrows point to regions of interest.}
\label{fig:illResult}
\end{figure}

The steady-state distribution of temperature and arsine mass fraction for case 3 (Tab.~\ref{tab:conditions}) with $K_{reac} = 222.8$~m.s$^{-1}$ and $A_{cracking} = 8.5\times 10^5$~m.s$^{-1}$ is illustrated in Fig.~\ref{fig:illResult}. It can be seen that since the flow inside the center port is heated to a temperature close to $1073.15$~K, the center port injects a hot flow at the substrate (indicated by the arrow). The contour of \ce{AsH3} (bottom) shows the effect of the flow curtain which avoids contamination of the adjacent chamber with arsine (left arrow). At the substrate, \ce{AsH3} is consumed, which results in a non-zero normal gradient (right arrow). Similar features can be observed for \ce{GaCl}.

\begin{figure}[ht!]
\centering
\includegraphics[width=0.45\textwidth,trim={0cm 0cm 0cm 0cm},clip]{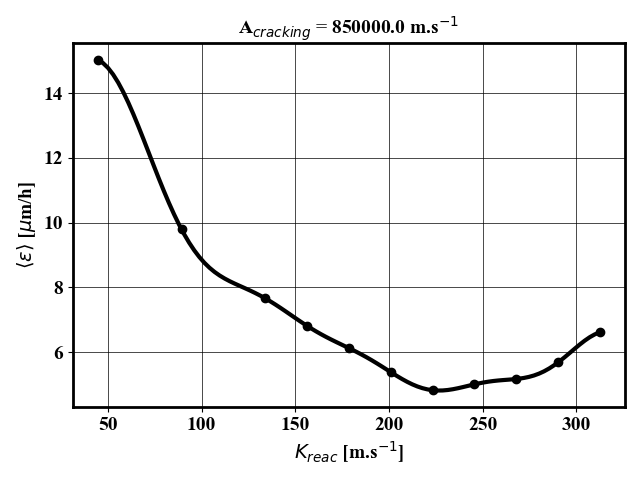}
\caption{Average growth rate prediction error across the seven cases investigated as a function of the kinetic constant $K_{reac}$ with $A_{cracking}=8.5\times10^5$~m.s$^{-1}$. Simulation results (\mythickcircle{black}{black}) and interpolated results with radial basis functions (\mythickline{black}) are displayed.}
\label{fig:calibrationResults}
\end{figure}

For each $(K_{reac}, A_{cracking})$ sample pair, the average error in growth rate $\langle \varepsilon \rangle$ over the seven cases considered is computed. The results are shown in Fig.~\ref{fig:calibrationResults}. It can be seen that there exists an optimal value of $K_{reac}$ which leads to the best match of the growth rates. In the particular case considered ($A_{cracking}=8.5\times10^5$~m.s$^{-1}$), $K_{reac} = 222.8$~m.s$^{-1}$ minimizes $\langle \varepsilon \rangle$. To improve the estimate of the value of $K_{reac}$ that minimizes the error, interpolation between data points is done with radial basis functions, where hyperparameters are optimized using the Scikit-Learn library \cite{scikit-learn}. An example of the interpolation result is shown in Fig.~\ref{fig:calibrationResults}.

\begin{figure}[ht!]
\centering
\includegraphics[width=0.45\textwidth,trim={0cm 0cm 0cm 0cm},clip]{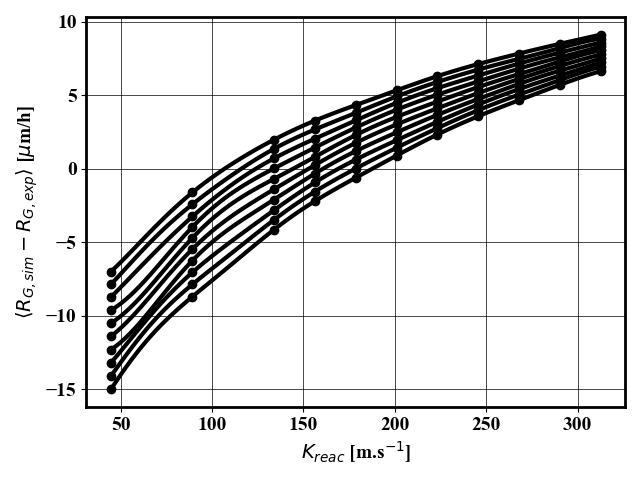}
\includegraphics[width=0.45\textwidth,trim={0cm 0cm 0cm 0cm},clip]{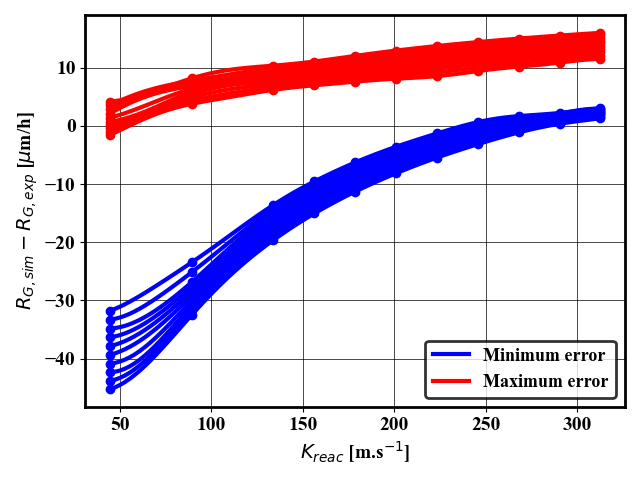}
\caption{Left: average difference between the rate of growth obtained from simulations $R_{G,sim}$ and the rate of growth obtained from experiments $R_{G,exp}$ plotted against $K_{reac}$ for all $A_{cracking}$. Right: minimum (\mythickline{blue}) and maximum (\mythickline{red}) difference of the rate of growth obtained from simulations $R_{G,sim}$ compared to rate of growth obtained from experiments $R_{G,exp}$ plotted against $K_{reac}$ for all $A_{cracking}$.}
\label{fig:globalMinimum}
\end{figure}

Over the range chosen $K_{reac}$, it can be observed from Fig.~\ref{fig:calibrationResults} that there exists a value that can serve, at least, as a global minimizer of the discrepancy with experimental results. It is shown here, that the local minimum is likely a global minimum. We first argue that the relationship between the rate of growth $R_{G}$ and the kinetic coefficient $K_{reac}$ is monotonic and increasing. Physically, this assertion is justified by the fact that increasing the kinetic rate should increase the rate of production of \ce{GaAs}. We also propose a numerical argument by plotting the difference between the rate of growth obtained in the simulations ($R_{G,sim}$) and in experiments ($R_{G,exp}$) for all $A_{cracking}$ and all $K_{reac}$ values. The results are shown in Fig.~\ref{fig:globalMinimum} (top). As expected, the rate of growth is monotonically increasing with $K_{reac}$ for all $A_{cracking}$ values. Next, we record the minimum and maximum error over all the cases for each $A_{cracking}$ and $K_{reac}$. Because of the monotonically increasing relation between $K_{reac}$ and $R_{G}$, if the maximum error $R_{G,sim}-R_{G,exp}$ at the lowest $K_{reac}$ value is negative, then $\forall K_{reac} < 45$~m.s$^{-1}$, $\langle \varepsilon \rangle$ would be larger than for $K_{reac} = 45$~m.s$^{-1}$ . In Fig.~\ref{fig:globalMinimum} (right), it can be seen that the maximum absolute value of the error is almost negative for all $A_{cracking}$ at $K_{reac}=45$~m.s$^{-1}$. The same argument can be invoked for the minimum error at $K_{reac}=310$~m.s$^{-1}$ which is positive. We therefore argue that the minimum shown is likely a global minimum.

\begin{figure}[ht!]
\centering
\includegraphics[width=0.45\textwidth,trim={0cm 0cm 0cm 0cm},clip]{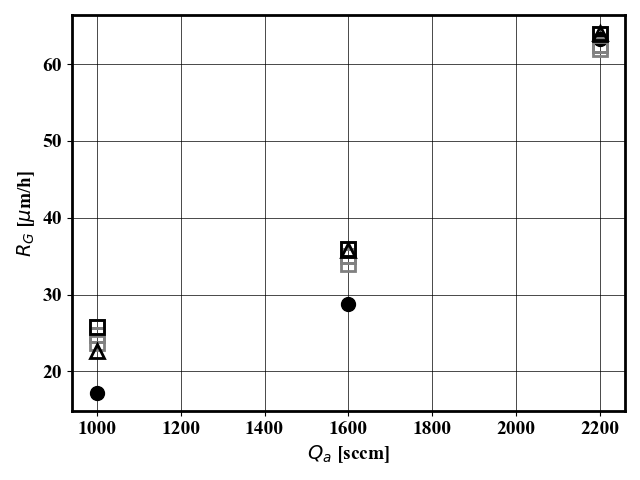}
\includegraphics[width=0.45\textwidth,trim={0cm 0cm 0cm 0cm},clip]{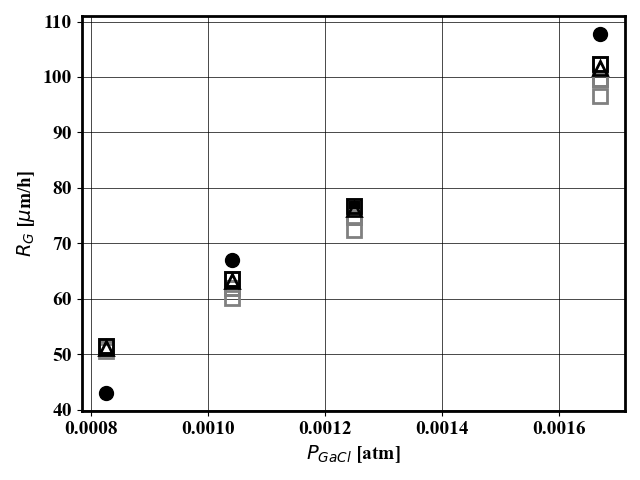}
\caption{Growth rate of \ce{GaAs} comparison between experimental data reported in \citet{schulte2018high} (\mythickcircle{black}{black}), the calibrated model used with the 0.5 million cells grid (\mythicksquare{black}{white}) and the calibrated model used with the 1 million cells grid (\mythicktriangle{black}{white}) for $A_{cracking}=8.5\times10^5$~m.s$^{-1}$. Results are overlayed with calibrated model obtained with the 0.5 million cells grid for all other $A_{cracking}$ values (\mythicksquare{gray}{white}). Top: Case 1-3 where growth rate is plotted against $Q_a$, the volume flow rate of \ce{H2} through the center port. Bottom: Case 4-7 where growth rate is plotted against $P_{GaCl}$, the partial pressure of \ce{GaCl}.}
\label{fig:comparisonExp}
\end{figure}

The growth rates computed with $K_{reac} = 222.8$~m.s$^{-1}$ and $A_{cracking}=8.5\times10^5$~m.s$^{-1}$ are plotted against experimental values in Fig.~\ref{fig:comparisonExp}. A reasonable agreement with experiments can be observed. Furthermore, the trends are consistent with the experiments and physical intuition. As the volume flow rate of \ce{H2}, $Q_a$ through the center port increases (left), less \ce{AsH3} cracks which leads to larger reaction rates. The exponential dependence of the amount of cracked \ce{AsH3} with the velocity (Eq.~\ref{eq:asH31d}) should result in a super-linear increase of growth rate with the flow rate of arsine through the center port. This is the trend observed in the experiments and the simulations. This observation is in sharp contrast with the linear increase of the growth rate with increasing concentration of \ce{GaCl} (bottom). To ensure that the calibrated $K_{reac}$ is only minimally affected by the numerical errors, the fine grid (1 million computational cells) is used to verify that the results are grid converged. The fine and coarse grids predict similar growth rates and therefore the baseline grid is sufficient for accurate predictions.

\begin{figure}[ht!]
\centering
\includegraphics[width=0.45\textwidth,trim={0cm 6cm 0cm 6cm},clip]{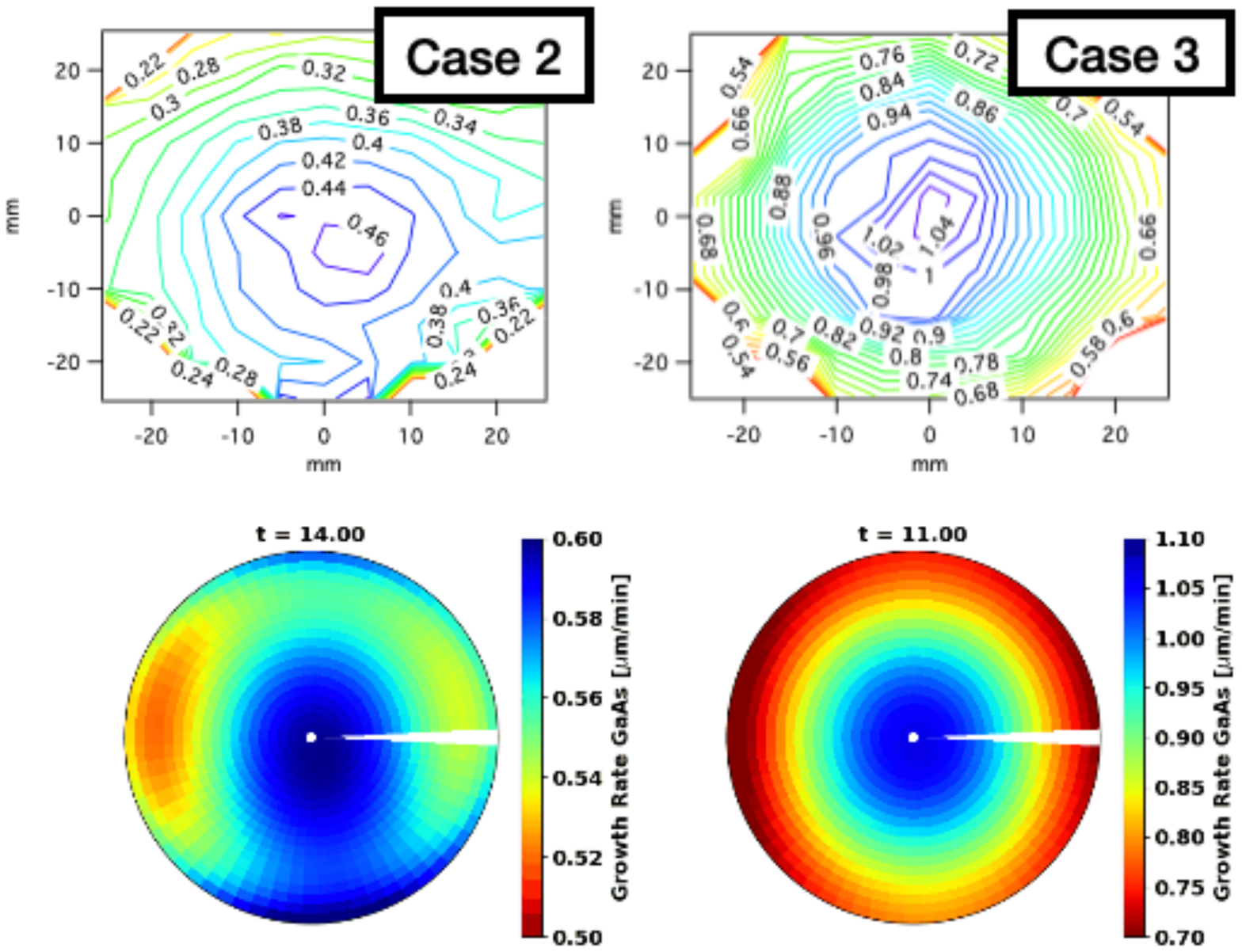}
\caption{Spatial distribution of growth rate at the substrate for case 2 (left) and case 3 (right) between experimental data (top) and simulation data (bottom) with $A_{cracking}=8.5\times10^5$~m.s$^{-1}$ and $K_{reac} = 222.8$~m.s$^{-1}$.}
\label{fig:spatialDistribution}
\end{figure}

The spatial distribution of the growth rate at the substrate is compared to experimental data in Fig.~\ref{fig:spatialDistribution}. A reasonable agreement can be observed for case 3 (strong flow through the center port). A breakdown of the axial symmetry can be observed for case 2 (weaker flow through the center port) both in simulation and experimental data. The apparent breakdown of symmetry typically appears at low center port velocity, such as the one of Case 2. The flow near the substrate is influenced by a vertical component that originates from the center port, and by a horizontal component, the crossflow that connects the two deposition chambers. The horizontal flow can be observed in Fig.~\ref{fig:illResult} (middle and bottom). When lower center port velocities are used, the relative effect of the horizontal crossflow may increase and disturb the boundary layer near the substrate, thereby leading to a breakdown of growth symmetry. Interestingly, the same symmetry breakdown seems to occur in the experimental results, albeit to a smaller extent. The discrepancy may be due to a slightly earlier center port jet breakdown in the simulations due to either numerical diffusion, or errors in the center jet exit boundary conditions. Finally, we note that in other experiments reported for the same reactor~\cite{schulte2019uniformity}, a similar breakdown of symmetry may appear at different conditions. This result suggests that the model can be used not only to assess the overall growth rate of the substrate but also its spatial uniformity, which is critical for the efficiency of solar cells.  

\begin{figure}[ht!]
\centering
\includegraphics[width=0.45\textwidth,trim={0cm 0cm 0cm 0cm},clip]{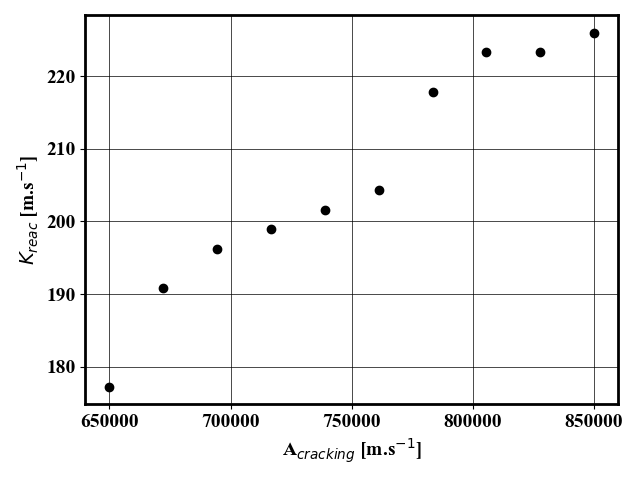}
\includegraphics[width=0.45\textwidth,trim={0cm 0cm 0cm 0cm},clip]{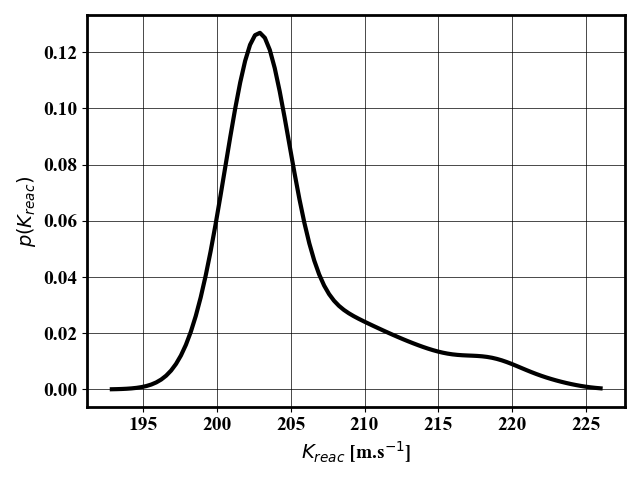}
\caption{Top: calibrated value of $K_{reac}$ for each value of $A_{cracking}$ considered. Bottom: constructed PDF of $K_{reac}$.}
\label{fig:responseSurface}
\end{figure}

As explained in Sec.~\ref{sec:calibrationMethod}, the calibration procedure is repeated for ten different values of $A_{cracking}$ which span the support of its posterior distribution obtained in Sec.~\ref{sec:cracking}. As a result, one obtains the response curve of the optimal $K_{reac}$ as a function of $A_{cracking}$, which is shown in Fig.~\ref{fig:responseSurface} (top). As $A_{cracking}$ increases, the amount of \ce{AsH3} that reaches the substrate decreases. Therefore, to keep constant the growth rates, the optimal $K_{reac}$ is a monotonically increasing function of $A_{cracking}$. This trend can be observed in Fig.~\ref{fig:responseSurface}. It can however be seen that the increase occurs by steps which is a consequence of the discretization of $A_{cracking}$. Given the dependence of $K_{reac}$ with $A_{cracking}$, one can sample $A_{cracking}$ using a Markov-chain Monte Carlo (MCMC) approach and construct a PDF of $K_{reac}$. The results are shown in Fig.~\ref{fig:responseSurface} (bottom). The standard deviation of the PDF of $K_{reac}$ can then be used to extract uncertainty bounds on the calibrated value of the kinetic parameter. We found $K_{reac} = 205 \pm 5$~m.s$^{-1}$

\section{Conclusions}
\label{sec:conclusions}

In this work, a comprehensive modeling procedure was developed for the simulation of growth rate via uncracked hydrides in a D-HVPE reactor. The \ce{AsH3} cracking model was adjusted for the geometry and a surrogate model for the cracking through the injectors was developed. Multiple simulations for the surface reactions at the substrate were used to calibrate a simple kinetic model for the reaction of uncracked arsine. The proposed model matches experimental growth rate values and spatial distributions for a range of typical HVPE operating conditions. 

As part of the calibration, an uncertainty quantification procedure was conducted to include uncertainty in the \ce{AsH3} cracking model parameters. While the influence of other uncertain parameters like numerical CFD errors was shown to be negligible, uncertainty due to the interpolation during the calibration procedure will be included as future work.

Since III-V solar cells typically require the growth of different layers, the same procedure will be applied to develop kinetic models that correspond to each one of the compounds. 
Finally, the validated kinetic model will be used in conjunction with a CFD software to design scaled-up reactors able to maximize growth rate while minimizing growth non-uniformity.

\begin{acknowledgments}
Fruitful discussions with Milo Parra-Alvarez and Olga Doronina are gratefully acknowledged. This work was authored by the National Renewable Energy Laboratory, operated by Alliance for Sustainable Energy, LLC, for the U.S. Department of Energy (DOE) under Contract No. DE-AC36-08GO28308. This work was supported by the Air Force Research Laboratory (IAG-19-02103). This research was performed using computational resources sponsored by the U.S. Department of Energy's Office of Energy Efficiency and Renewable Energy and located at the National Renewable Energy Laboratory. The views expressed in the article do not necessarily represent the views of the DOE or the U.S. Government. The U.S. Government retains and the publisher, by accepting the article for publication, acknowledges that the U.S. Government retains a nonexclusive, paid-up, irrevocable, worldwide license to publish or reproduce the published form of this work or allow others to do so, for U.S. Government purposes. 
\end{acknowledgments}

\section*{Data Availability}

The data that support the findings of this study are available from the corresponding author upon reasonable request.

\appendix

\section{Fluid solver}
\label{app:fluidSolver}

A variable density low-Mach number solver based on the OpenFOAM framework is used \cite{hassanaly2018minimally} to model the flow inside the reactor. The solver has been successfully applied in several other studies \cite{hassanaly2020data,tang2021large} including configuration with reactive phase transformations such as the one considered here \cite{koo2017large}. The low-Mach number assumption allows for the use larger time-steps on the order of fluid convection time-scales($\sim$ $1$ ms) in all the simulations, without the need to resolve acoustic time-scales. Thermodynamics variables (specific heat, enthalpy) and transport properties (viscosity, thermal conductivity, species diffusivity) used in the model are obtained from open source library, \textit{PelePhysics}, designed for complex combustion simulations \cite{sitaraman2021adaptive}.

The momentum transport equation given by:

\begin{equation}
    \frac{\partial \rho \boldsymbol{u}}{ \partial t} + \nabla \cdot (\rho \boldsymbol{u} \boldsymbol{u}) = -\nabla p + \nabla \cdot \boldsymbol{\overline{\tau}},
\end{equation}

is solved, where the pressure $p$ is computed to ensure mass conservation and $\boldsymbol{\overline{\tau}}$ is the stress tensor.

The gas-phase species are transported using the equation 

\begin{equation}
    \frac{\partial \rho Y_i}{ \partial t} + \nabla \cdot (\rho \boldsymbol{u} Y_i) = \nabla \cdot (\rho D_i \nabla Y_i) + \dot{\omega_i},
\end{equation}

where $Y_i~\in~\{ Y_{\ce{AsH3}}, Y_{\ce{As4}}, Y_{\ce{GaCl}}, Y_{\ce{HCl}} \}$ are species mass fractions, $\rho$ is the gas mixture density, $\boldsymbol{u}$ is the flow velocity and $D_i$ are the individual species diffusivities. The mass fraction of the carrier specie $Y_{\ce{H2}}$ is obtained by enforcing the unity mass fraction sum constraint.

The energy equation is solved in terms of a global temperature among all species given by: 

\begin{equation}
    \label{eq:energyTransport}
    \frac{\partial \rho T}{ \partial t} + \nabla \cdot (\rho \boldsymbol{u} T) = \nabla \cdot (\lambda \nabla T) - \rho (\nabla T) \cdot \sum_{i=1}^N Cp_i Y_i \boldsymbol{V_i} + \dot{\omega_{HR}},
\end{equation}

where $T$ denotes temperature, $\lambda$ is the thermal conductivity, $Cp_i$ are the species heat capacities, $\boldsymbol{V_i}$ are the species diffusion velocities, $N$ is the number of species, and $\dot{\omega_{HR}}$ denotes the heat release rate. 

A Smagorinsky \cite{smagorinsky1963general} turbulence model is used and is mostly active in the shear layer near the center-port exit, while the flow near the substrate is nearly laminar. For the scalar turbulence, a turbulent Schmidt number $Sc_t = 0.72$ and a turbulent Prandtl number $Pr = 0.7$ were used \cite{tang2019comprehensive}. The turbulence model is useful for stabilization purposes in the shear layers and all the results in this work are obtained when the reactor reaches steady-state. 

\section{Numerical implementation of finite-rate surface reactions}
\label{app:numericsSurfaceReaction}
The surface reactions are implemented in a fashion similar to \citet{maestri2013coupling} by imposing a non-zero flux of species at the walls. Formally, the reaction source term for species $i$ is written as 

\begin{equation}
    \dot{\omega_i} = \nabla \cdot \Omega_i, 
\end{equation}

where $\Omega_i$ is a field defined at the cell faces which is null everywhere in the domain, except at the walls where surface reactions occur. $\Omega_i$ is computed using the kinetic model and by assuming the surface concentration of the species equals that of the volumetric concentration at the computational cell immediately adjacent to the face. The same procedure is used to define the heat release from the walls. The surface reactions are coupled with the scalar transport equation using a Strang splitting method \cite{strang1968construction}, except when infinitely fast chemistry is used (See App.~\ref{app:infiniteChemistry}).

The surface reactions in this work involve a conversion of species in the gas phase to solid phase products at the catalytic surfaces. Therefore, mass may leave or enter the gas phase at these surfaces. Similar to other cases where a phase change occurs like with spray evaporation \cite{heye2013probability} or soot formation \cite{chong2018large}, the pressure equation needs to be augmented with a source term for mass. The mass source term $\dot{m}$ is simply computed as 

\begin{equation}
    \label{eq:pressure}
    \dot{m} = \sum_{i=1}^N \dot{\omega_i},
\end{equation}

where $N$ is the number of species.

\section{Thermodynamic and transport parameters}
\label{app:thermoParam}
The thermodynamic parameters (enthalpy and calorific capacity) are obtained from polynomial fits that relate them to the local temperature. The coefficients of the polynomial fits are obtained from ~\citet{oehlschlaeger2009autoignition}. Missing properties of \ce{GaCl} were substituted with the properties of \ce{GaH}. Note that although the polynomial coefficient for the calorific capacity and enthalpy were substituted, the molar weight (used to compute density) was not modified.
Similarly, the transport parameters (species diffusivity, heat conductivity, dynamic viscosity) are computed at every grid point using the species concentrations and temperature. The transport parameters are computed using the following molecular properties: their Lennard Jones potential well depth, collision diameter, dipole moment, polarizability, and their rotational relaxation collision number. The molecular properties are obtained from \citet{kee1999transport}. Again, since the properties of some species were not available, they had to be substituted with other ones. In particular, the molecular properties of \ce{GaCl} and \ce{As4} were substituted with that of \ce{GaH} and \ce{As2}.

\section{Cracking model validation and center port surrogate model}
\label{app:surrogateModelCenterPort}

The two assumptions used to derive the cracking model namely the 1D advection-reaction approximation and the sole dependence of density on temperature are addressed. The model is compared to three-dimensional reacting flow simulations. The validation of these assumptions will further bolster the correction applied to the cracking kinetic coefficients.    

The 1D transport of enthalpy in a cylindrical tube neglecting axial diffusion can be approximated as 
\begin{equation}
    \frac{d}{dx} (\rho U Cp T) \pi R^2 dx = h (T_w(x) - T(x)) 2 \pi R dx,
\end{equation}

where $\rho$ is the local density, $Cp$ is the local heat capacity, $T(x)$ is the local temperature, $T_w(x)$ is the wall temperature and $R$ is the tube radius and $h$ is the convective heat transfer coefficient. The convective heat transfer coefficient for a fully developed pipe flow can be obtained from Nusselt number $Nu=3.66$ \cite{bergman2011fundamentals}. The heat transfer coefficient can be written as $h \approx Nu \lambda/(R)$ where $\lambda$ is the thermal conductivity. Assuming a near-constant heat capacity and neglecting the addition of mass due to surface reactions

\begin{equation}
    \label{eq:t1d}
    \frac{d T}{dx} = \frac{2 h}{R \rho_0 U_0 Cp_0} (T_w(x) - T(x)),
\end{equation}

where $\rho_0$, $U_0$, and $Cp_0$ are the fluid properties at the pipe inlet. The same approach for the transport of mass fraction of \ce{AsH3} leads to

\begin{multline}
  \label{eq:asH31d}
    \frac{d Y_{\ce{AsH3}}}{dx} = \frac{-2 Y_{\ce{AsH3}}}{R \rho_0 U_0} A_{cracking}\,\,e^{\frac{-Ea_{cracking}}{R_u T_w(x)}} \frac{\rho_0 T_0}{T(x)},
\end{multline}

where $\rho(x)$ was assumed to be solely a function of temperature. Given a 1D discretization, Eqs.~\ref{eq:t1d} and \ref{eq:asH31d} can be solved with an exponential type of integrator such as a Rosenbrock integrator. 

\begin{figure}[ht!]
\centering
\includegraphics[width=0.15\textwidth,trim={5cm 5cm 5cm 5cm},clip]{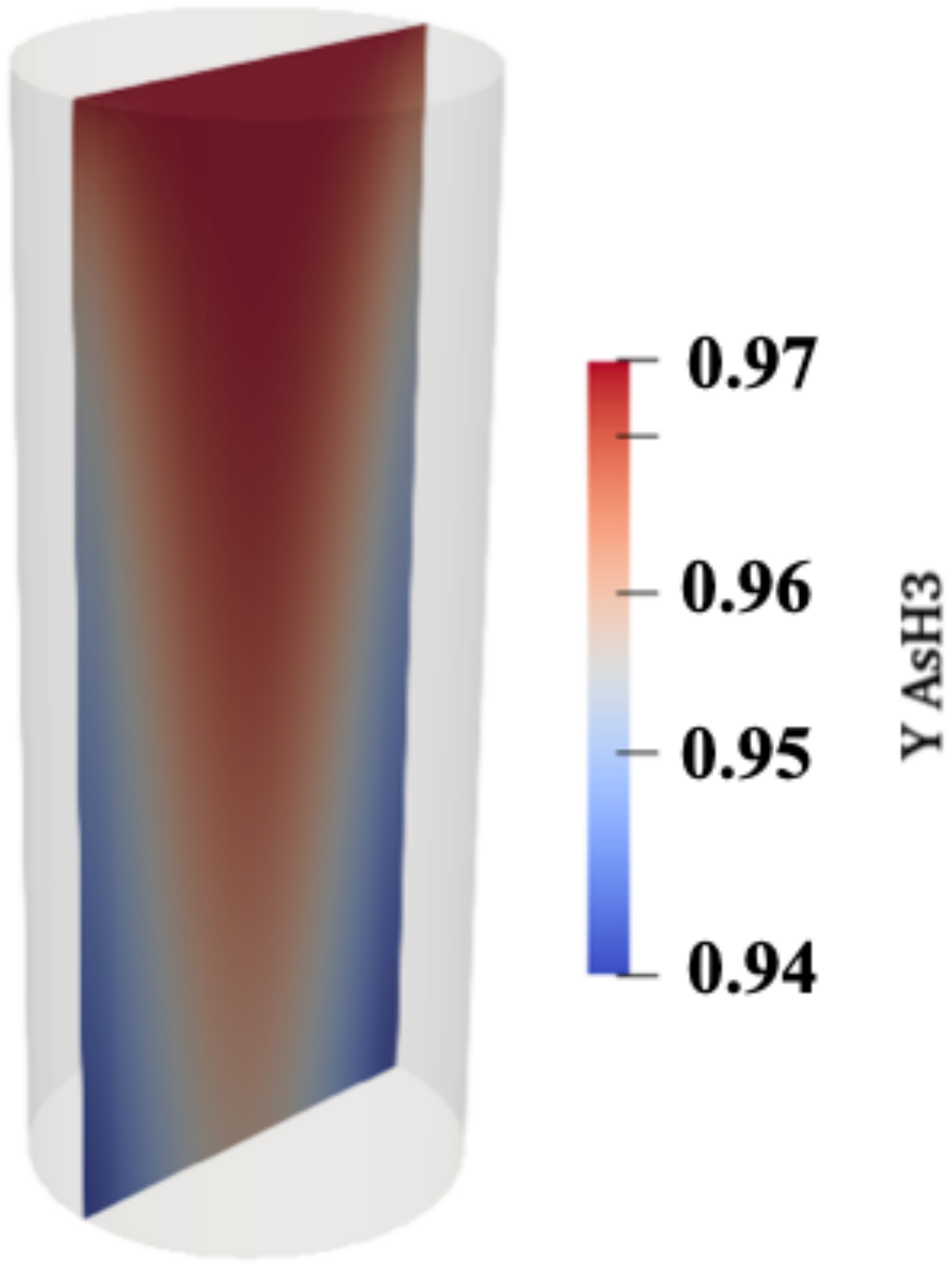}
\includegraphics[width=0.27\textwidth]{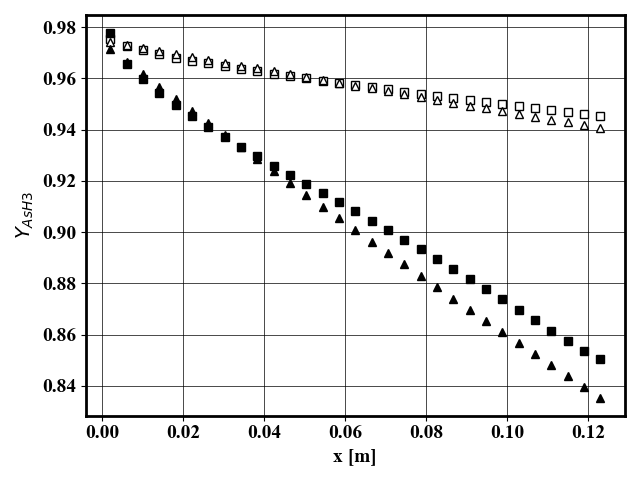}
\caption{Left: contour of mass fraction of \ce{AsH3} in a cylinder that replicates the experiments \cite{denbaars1986homogeneous}. Flow is from top to bottom. Right: surface averaged longitudinal profile of mass fraction of \ce{AsH3} for a wall temperature of $723.15$~K for the 3D model (\mythicksquare{black}{white}) and the 1D model (\mythicktriangle{black}{white}) and for a wall temperature of $773.15$~K for the 3D model (\mythicksquare{black}{black}) and the 1D model (\mythicktriangle{black}{black}).}
\label{fig:denbaarCyl}
\end{figure}

In Fig.~\ref{fig:denbaarCyl} the results for a 3D reacting flow simulation are shown for conditions similar to the ones of \citet{denbaars1986homogeneous}. The wall temperature is held fixed at $723.15$~K and $773.15$~K, the cylinder radius is $25$~mm, its length is $125$~mm, the pressure is $1$~atm, and the inlet is made of 50\% \ce{AsH3} and \ce{H2} by volume. The mesh size is $1.5$~mm radially and $4$~mm axially.

The profile of $Y_{\ce{AsH3}}$ exhibits variation within a cross-section due to the relatively larger diameter. Nevertheless, the 1D model successfully approximates the longitudinal profile of $Y_{\ce{AsH3}}$. In turn, it suggests that the correction for the kinetic rate for arsine cracking is valid.

\begin{figure}[ht!]
\centering
\includegraphics[width=0.15\textwidth,trim={5cm 5cm 5cm 5cm},clip]{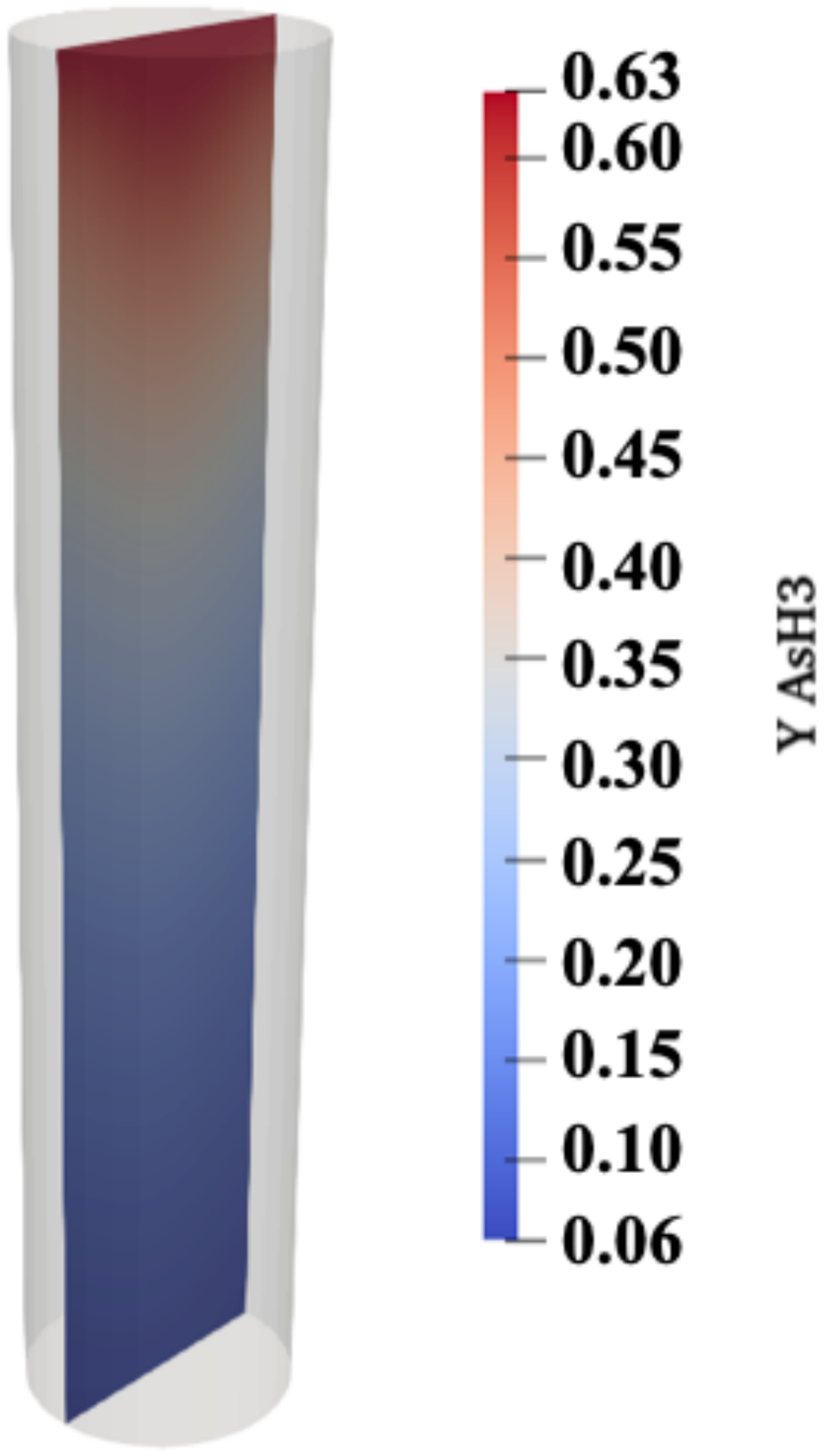}
\includegraphics[width=0.32\textwidth]{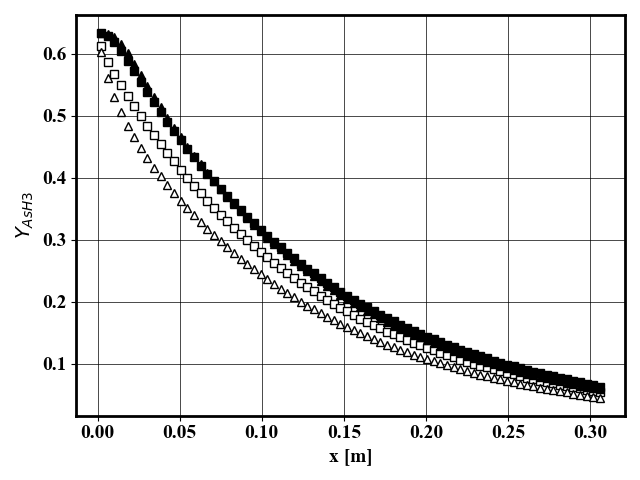}
\caption{Left: contour of mass fraction of \ce{AsH3} in a cylinder that replicates the center port. The longitudinal direction was rescaled by a factor 0.032 for plotting purposes. Flow is from top to bottom. Right: surface averaged longitudinal profile of mass fraction of \ce{AsH3} for an outer-wall temperature of $1023.15$~K without taking into account the heat transfer through the quartz for the 3D model (\mythicksquare{black}{white}) and the 1D model (\mythicktriangle{black}{white}), and by taking into account the heat transfer through the quartz for the 3D model (\mythicksquare{black}{black}) and the 1D model (\mythicktriangle{black}{black}).}
\label{fig:centerPortCyl}
\end{figure}

The same procedure can be applied to the center port of the HVPE reactor studied in this work. In this case, the cylinder has a $2$~mm radius and its length is $308$~mm. The center port is even more suited to the 1D approximation given its larger length to radius ratio. The wall temperature is held fixed at $1023.15$~K, and the pressure is $0.829$~atm. The mesh size is $0.12$~mm radially and $4$~mm axially. The results are shown in Fig.~\ref{fig:centerPortCyl}. Since the aspect ratio of the cylinder is smaller than that of \citet{denbaars1986homogeneous}, the profile of \ce{AsH3} is almost one dimensional (left). In turn, the 1D model accurately reproduces the longitudinal profile of \ce{AsH3}. 

In the experiments reported in \citet{schulte2018high}, the center port has an inner radius $R_{in}=2$~mm and an outer radius of $R_{out}=4.5$~mm. One can also incorporate the conjugate heat transfer between the quartz annulus and the inner flow. The inner-wall temperature can be obtained by writing a balance of heat flux at the inner-wall boundary 

\begin{equation}
    h (T(x) - T_{iw}(x)) + \frac{\lambda_{quartz}}{\Delta R} (T_{w}(x)-T_{iw}(x)) = 0,
\end{equation}

where $\lambda_{quartz}$ is the heat conductivity of the quartz, $T_{iw}$ is the inner-wall temperature, $T_{w}$ is the outer-wall temperature and $\Delta R= R_{out} - R_{in}$. An algebraic expression can be obtained for $T_{iw}$, and Eq.~\ref{eq:t1d} becomes

\begin{equation}
    \label{eq:t1dCHT}
    \frac{d T}{dx} = \frac{2 h}{R \rho_0 U_0 Cp_0 \left(1 + \frac{h \Delta R}{ \lambda_{quartz}}\right)} (T_w(x) - T(x)),
\end{equation}

and Eq.~\ref{eq:asH31d} becomes
\begin{equation}
    \label{eq:asH31dCHT}
    \frac{d Y_{\ce{AsH3}}}{dx} = \frac{-2 Y_{\ce{AsH3}}}{R \rho_0 U_0} A_{cracking} e^{-Ea_{cracking}/(R_u T_{iw}(x))} \frac{\rho_0 T_0}{T(x)},
\end{equation}

where $T_{iw}(x) = T(x) + \frac{T_w(x)-T(x)}{1 + h\Delta R/\lambda_{quartz}}$.

The 1D model is compared to a 3D model run with CFD, where the heat transfer through the solid is written assuming constant thermodynamic and transport properties 

\begin{equation}
    \label{eq:solidHeat}
    \frac{\partial}{\partial t} (\rho_{quartz} T) = \frac{1}{Cp_{quartz}} \nabla \cdot (\lambda_{quartz} \nabla T),
\end{equation}

where $\rho_{quartz}$ is the density of the solid assumed to be equal to $2200$~kg.m$^{-3}$, $Cp_{quartz}$ is the heat capacity of the solid set at $705$~J.kg$^{-1}$.K$^{-1}$, and $\lambda_{quartz}$ is equal to $1.46$~W.m$^{-1}$.K$^{-1}$. The coupling between the solid and fluid equations is performed via outer-Picard iterations \cite{hassanaly2018minimally}. The thermal coupling at every face of the inner-wall boundary is done by dynamically setting the inner-wall temperature to 

\begin{equation}
    T_{iw} = \frac{T_s \frac{\lambda_{s}}{\Delta s} + T_f \frac{\lambda_{fluid}}{\Delta f}}{\frac{\lambda_{s}}{\Delta s} + \frac{\lambda_{fluid}}{\Delta f}},
\end{equation}

where $T_s$ (respectively $T_f$) is the temperature in the solid (respectively fluid) domain that is radially adjacent to the inner-wall face, $\lambda_{s}$ (respectively $\lambda_f$) is the heat transfer coefficient at that same location, $\Delta s$ (respectively $\Delta_f$) is the distance between that location and the inner-wall face center. The choice of the inner-wall temperature boundary condition equates both the temperature and the heat flux at the inner-wall interface. The mesh size is the same as the one used without the conjugate heat transfer, and the results are shown in Fig.~\ref{fig:centerPortCyl} (right) for an outer-wall temperature of $1023.15$~K. The 1D model and the 3D model reasonably agree on the longitudinal profile of \ce{AsH3}. Because the inner-wall temperature is lower than without the conjugate heat transfer, less \ce{AsH3} cracks through the center port. Similar to the results shown in Fig.~\ref{fig:denbaarCyl}, since the 1D model neglects radial diffusion, it overestimates the amount of \ce{AsH3} immediately adjacent to the walls, thereby slightly overestimating the amount of \ce{AsH3} cracked. 

Overall, the 1D model is a reasonable representation at lower computational costs for modeling transport and chemistry within the center port and it is proposed to be used as a surrogate for the flow inside the center port to obtain the boundary conditions at the exit of the injector. The boundary conditions for velocity can be obtained from mass conservation assuming that it solely depends on the density at the exit of the tube which, in turn, is assumed to be a sole function of temperature. The mass fraction of \ce{As4} at the center port exit can be computed by integrating Eq.\ \ref{eq:asH31dCHT} over the length of the tube.

In the CFD model, the surrogate model is implemented by first collecting the value of temperature (transported according to Eq.~\ref{eq:energyTransport}) at the outer-walls of the center port, at every timestep. The temperature values are averaged azimuthally to construct a one-dimensional representation of the outer-wall temperature of the center port. The one-dimensional wall temperature profile is then directly used in the advection-reaction model of the center port. The boundary conditions at the injector exit are subsequently evaluated by integrating the 1D ordinary differential equation systems (Eqs.\ \ref{eq:t1dCHT} and \ref{eq:asH31dCHT}) for the scalars and temperature. This procedure allows incorporating longitudinal gradients of temperature along the center-port walls into the center-port exit boundary condition. Although the method is repeated at every timestep, the overhead computational cost is negligible compared to momentum and scalar transport.



\section{Effect of prior on the Bayesian calibration}
\label{app:bayesianPrior}
\begin{figure}[ht!]
\centering
\includegraphics[width=0.45\textwidth,trim={0cm 0cm 0cm 0cm},clip]{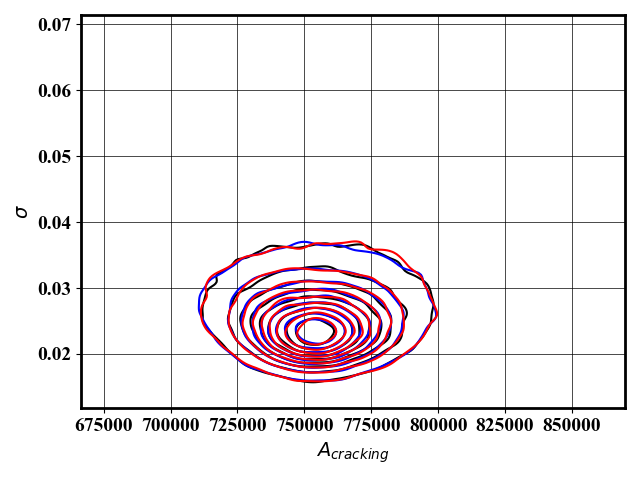}
\caption{Black line: Two-dimensional posterior PDF for prior used  ($\mathcal{U}(10^4,2\times 10^7)$ for $A_{cracking}$ and $\mathcal{U}(0,10)$ for $\sigma$). Blue line: Two-dimensional posterior PDF for narrow prior  ($\mathcal{U}(10^5,10^6)$ for $A_{cracking}$ and $\mathcal{U}(0,1)$ for $\sigma$). Red line: Two-dimensional posterior PDF for wide prior ($\mathcal{U}(1,2\times10^{11})$ for $A_{cracking}$ and $\mathcal{U}(0,100)$ for $\sigma$).}
\label{fig:priorEffect}
\end{figure}

In Bayesian analysis, the posterior distribution depends on the prior, which in turn may have an effect on the PDF of the calibrated $K_{reac}$. In the results reported in the paper, the prior was chosen to be $\mathcal{U}(10^4,2\times10^7)$ for $A_{cracking}$ and $\mathcal{U}(0,10)$ for $\sigma$ . In Fig.~\ref{fig:priorEffect}, the results are reported with a narrow prior $\mathcal{U}(10^5,10^6)$ for $A_{cracking}$ and $\mathcal{U}(0,1)$ for $\sigma$; and a wide prior $\mathcal{U}(1,2\times 10^{11})$ for $A_{cracking}$ and $\mathcal{U}(0,100)$ for $\sigma$. As can be observed, the prior has little effect on the posterior distribution of $A_{cracking}$.

\section{Infinitely fast kinetic growth model}
\label{app:infiniteChemistry}

An infinitely fast kinetic model is implemented following the hypothesis that the reactor operates in a transport-limited regime. A mass flux boundary condition is imposed for the species transport equations at the substrate using the reactant composition at adjacent computational cells. Molar stoichiometry is used to determine excess and deficient reactants at the surface and surface gradients are subsequently set to match the desired consumption rate.

The growth rate results are reported for Case 2 and Case 3 in Tab.~\ref{tab:infinitelyFast}. While the order of magnitude of the growth rates obtained with the infinitely fast chemistry is reasonable and has the correct trend compared to experiments, it leads to growth rates about 2 times larger (third row) than the ones observed in the experiments (second row). We conclude that a simple transport-limited description of the reaction cannot accurately explain the growth rate of \ce{GaAs} via uncracked hydrides.


\begin{table}[ht!]
\setlength{\abovecaptionskip}{0pt}
\setlength{\belowcaptionskip}{5pt}
\caption{Growth rates of Case 2 and 3 obtained with infinitely fast chemistry.}
\label{tab:infinitelyFast}
\begin{ruledtabular}
\begin{tabular}{|l|c|c|}
\hline
 &
 $R_G$ [$\mu m/h$] for Case 2  &
 $R_G$ [$\mu m/h$] for Case 3   \\ \hline
\citet{schulte2018high} &
28.8 &
63.2 \\ \hline
Infinitely fast  &
59.1 &
125.1

\end{tabular}
\end{ruledtabular}
\end{table}

\nocite{*} 
%


\begin{thebibliography}{34}%
\makeatletter
\providecommand \@ifxundefined [1]{%
 \@ifx{#1\undefined}
}%
\providecommand \@ifnum [1]{%
 \ifnum #1\expandafter \@firstoftwo
 \else \expandafter \@secondoftwo
 \fi
}%
\providecommand \@ifx [1]{%
 \ifx #1\expandafter \@firstoftwo
 \else \expandafter \@secondoftwo
 \fi
}%
\providecommand \natexlab [1]{#1}%
\providecommand \enquote  [1]{``#1''}%
\providecommand \bibnamefont  [1]{#1}%
\providecommand \bibfnamefont [1]{#1}%
\providecommand \citenamefont [1]{#1}%
\providecommand \href@noop [0]{\@secondoftwo}%
\providecommand \href [0]{\begingroup \@sanitize@url \@href}%
\providecommand \@href[1]{\@@startlink{#1}\@@href}%
\providecommand \@@href[1]{\endgroup#1\@@endlink}%
\providecommand \@sanitize@url [0]{\catcode `\\12\catcode `\$12\catcode
  `\&12\catcode `\#12\catcode `\^12\catcode `\_12\catcode `\%12\relax}%
\providecommand \@@startlink[1]{}%
\providecommand \@@endlink[0]{}%
\providecommand \url  [0]{\begingroup\@sanitize@url \@url }%
\providecommand \@url [1]{\endgroup\@href {#1}{\urlprefix }}%
\providecommand \urlprefix  [0]{URL }%
\providecommand \Eprint [0]{\href }%
\providecommand \doibase [0]{http://dx.doi.org/}%
\providecommand \selectlanguage [0]{\@gobble}%
\providecommand \bibinfo  [0]{\@secondoftwo}%
\providecommand \bibfield  [0]{\@secondoftwo}%
\providecommand \translation [1]{[#1]}%
\providecommand \BibitemOpen [0]{}%
\providecommand \bibitemStop [0]{}%
\providecommand \bibitemNoStop [0]{.\EOS\space}%
\providecommand \EOS [0]{\spacefactor3000\relax}%
\providecommand \BibitemShut  [1]{\csname bibitem#1\endcsname}%
\let\auto@bib@innerbib\@empty
\bibitem [{\citenamefont {King}\ \emph {et~al.}(2006)\citenamefont {King},
  \citenamefont {Fetzer}, \citenamefont {Law}, \citenamefont {Edmondson},
  \citenamefont {Yoon}, \citenamefont {Kinsey}, \citenamefont {Krut},
  \citenamefont {Ermer}, \citenamefont {Hebert}, \citenamefont {Cavicchi} \emph
  {et~al.}}]{king2006advanced}%
  \BibitemOpen
  \bibfield  {author} {\bibinfo {author} {\bibfnamefont {R.~R.}\ \bibnamefont
  {King}}, \bibinfo {author} {\bibfnamefont {C.~M.}\ \bibnamefont {Fetzer}},
  \bibinfo {author} {\bibfnamefont {D.~C.}\ \bibnamefont {Law}}, \bibinfo
  {author} {\bibfnamefont {K.~M.}\ \bibnamefont {Edmondson}}, \bibinfo {author}
  {\bibfnamefont {H.}~\bibnamefont {Yoon}}, \bibinfo {author} {\bibfnamefont
  {G.~S.}\ \bibnamefont {Kinsey}}, \bibinfo {author} {\bibfnamefont {D.~D.}\
  \bibnamefont {Krut}}, \bibinfo {author} {\bibfnamefont {J.~H.}\ \bibnamefont
  {Ermer}}, \bibinfo {author} {\bibfnamefont {P.}~\bibnamefont {Hebert}},
  \bibinfo {author} {\bibfnamefont {B.~T.}\ \bibnamefont {Cavicchi}},  \emph
  {et~al.},\ }\bibfield  {title} {\enquote {\bibinfo {title} {{Advanced III-V
  multijunction cells for space}},}\ }in\ \href@noop {} {\emph {\bibinfo
  {booktitle} {2006 IEEE 4th World Conference on Photovoltaic Energy
  Conference}}},\ Vol.~\bibinfo {volume} {2}\ (\bibinfo {organization} {IEEE},\
  \bibinfo {year} {2006})\ pp.\ \bibinfo {pages} {1757--1762}\BibitemShut
  {NoStop}%
\bibitem [{\citenamefont {Hoheisel}, \citenamefont {Philipps},\ and\
  \citenamefont {Bett}(2010)}]{hoheisel2010long}%
  \BibitemOpen
  \bibfield  {author} {\bibinfo {author} {\bibfnamefont {R.}~\bibnamefont
  {Hoheisel}}, \bibinfo {author} {\bibfnamefont {S.}~\bibnamefont {Philipps}},
  \ and\ \bibinfo {author} {\bibfnamefont {A.}~\bibnamefont {Bett}},\
  }\bibfield  {title} {\enquote {\bibinfo {title} {{Long-term energy production
  of III--V triple-junction solar cells on the Martian surface}},}\ }\href@noop
  {} {\bibfield  {journal} {\bibinfo  {journal} {Progress in Photovoltaics:
  Research and Applications}\ }\textbf {\bibinfo {volume} {18}},\ \bibinfo
  {pages} {90--99} (\bibinfo {year} {2010})}\BibitemShut {NoStop}%
\bibitem [{\citenamefont {Simon}\ \emph {et~al.}(2019)\citenamefont {Simon},
  \citenamefont {Schulte}, \citenamefont {Horowitz}, \citenamefont {Remo},
  \citenamefont {Young},\ and\ \citenamefont {Ptak}}]{simon2019iii}%
  \BibitemOpen
  \bibfield  {author} {\bibinfo {author} {\bibfnamefont {J.}~\bibnamefont
  {Simon}}, \bibinfo {author} {\bibfnamefont {K.~L.}\ \bibnamefont {Schulte}},
  \bibinfo {author} {\bibfnamefont {K.~A.}\ \bibnamefont {Horowitz}}, \bibinfo
  {author} {\bibfnamefont {T.}~\bibnamefont {Remo}}, \bibinfo {author}
  {\bibfnamefont {D.~L.}\ \bibnamefont {Young}}, \ and\ \bibinfo {author}
  {\bibfnamefont {A.~J.}\ \bibnamefont {Ptak}},\ }\bibfield  {title} {\enquote
  {\bibinfo {title} {{III-V-based optoelectronics with low-cost dynamic hydride
  vapor phase epitaxy}},}\ }\href@noop {} {\bibfield  {journal} {\bibinfo
  {journal} {Crystals}\ }\textbf {\bibinfo {volume} {9}},\ \bibinfo {pages} {3}
  (\bibinfo {year} {2019})}\BibitemShut {NoStop}%
\bibitem [{\citenamefont {McClure}\ \emph {et~al.}(2020)\citenamefont
  {McClure}, \citenamefont {Schulte}, \citenamefont {Simon}, \citenamefont
  {Metaferia},\ and\ \citenamefont {Ptak}}]{mcclure2020gaas}%
  \BibitemOpen
  \bibfield  {author} {\bibinfo {author} {\bibfnamefont {E.~L.}\ \bibnamefont
  {McClure}}, \bibinfo {author} {\bibfnamefont {K.~L.}\ \bibnamefont
  {Schulte}}, \bibinfo {author} {\bibfnamefont {J.}~\bibnamefont {Simon}},
  \bibinfo {author} {\bibfnamefont {W.}~\bibnamefont {Metaferia}}, \ and\
  \bibinfo {author} {\bibfnamefont {A.~J.}\ \bibnamefont {Ptak}},\ }\bibfield
  {title} {\enquote {\bibinfo {title} {Gaas growth rates of 528 $\mu$ m/h using
  dynamic-hydride vapor phase epitaxy with a nitrogen carrier gas},}\
  }\href@noop {} {\bibfield  {journal} {\bibinfo  {journal} {Applied Physics
  Letters}\ }\textbf {\bibinfo {volume} {116}},\ \bibinfo {pages} {182102}
  (\bibinfo {year} {2020})}\BibitemShut {NoStop}%
\bibitem [{\citenamefont {Metaferia}\ \emph {et~al.}(2019)\citenamefont
  {Metaferia}, \citenamefont {Schulte}, \citenamefont {Simon}, \citenamefont
  {Johnston},\ and\ \citenamefont {Ptak}}]{metaferia2019gallium}%
  \BibitemOpen
  \bibfield  {author} {\bibinfo {author} {\bibfnamefont {W.}~\bibnamefont
  {Metaferia}}, \bibinfo {author} {\bibfnamefont {K.~L.}\ \bibnamefont
  {Schulte}}, \bibinfo {author} {\bibfnamefont {J.}~\bibnamefont {Simon}},
  \bibinfo {author} {\bibfnamefont {S.}~\bibnamefont {Johnston}}, \ and\
  \bibinfo {author} {\bibfnamefont {A.~J.}\ \bibnamefont {Ptak}},\ }\bibfield
  {title} {\enquote {\bibinfo {title} {Gallium arsenide solar cells grown at
  rates exceeding 300 $\mu$m/h by hydride vapor phase epitaxy},}\ }\href@noop
  {} {\bibfield  {journal} {\bibinfo  {journal} {Nature communications}\
  }\textbf {\bibinfo {volume} {10}},\ \bibinfo {pages} {1--8} (\bibinfo {year}
  {2019})}\BibitemShut {NoStop}%
\bibitem [{\citenamefont {Horowitz}\ \emph {et~al.}(2018)\citenamefont
  {Horowitz}, \citenamefont {Remo}, \citenamefont {Smith},\ and\ \citenamefont
  {Ptak}}]{horowitz2018techno}%
  \BibitemOpen
  \bibfield  {author} {\bibinfo {author} {\bibfnamefont {K.~A.}\ \bibnamefont
  {Horowitz}}, \bibinfo {author} {\bibfnamefont {T.~W.}\ \bibnamefont {Remo}},
  \bibinfo {author} {\bibfnamefont {B.}~\bibnamefont {Smith}}, \ and\ \bibinfo
  {author} {\bibfnamefont {A.~J.}\ \bibnamefont {Ptak}},\ }\href@noop {}
  {\enquote {\bibinfo {title} {{A techno-economic analysis and cost reduction
  roadmap for III-V solar cells}},}\ }\bibinfo {type} {Tech. Rep.}\ (\bibinfo
  {institution} {National Renewable Energy Lab.(NREL), Golden, CO (United
  States)},\ \bibinfo {year} {2018})\BibitemShut {NoStop}%
\bibitem [{\citenamefont {Schulte}\ \emph {et~al.}(2018)\citenamefont
  {Schulte}, \citenamefont {Braun}, \citenamefont {Simon},\ and\ \citenamefont
  {Ptak}}]{schulte2018high}%
  \BibitemOpen
  \bibfield  {author} {\bibinfo {author} {\bibfnamefont {K.~L.}\ \bibnamefont
  {Schulte}}, \bibinfo {author} {\bibfnamefont {A.}~\bibnamefont {Braun}},
  \bibinfo {author} {\bibfnamefont {J.}~\bibnamefont {Simon}}, \ and\ \bibinfo
  {author} {\bibfnamefont {A.~J.}\ \bibnamefont {Ptak}},\ }\bibfield  {title}
  {\enquote {\bibinfo {title} {High growth rate hydride vapor phase epitaxy at
  low temperature through use of uncracked hydrides},}\ }\href@noop {}
  {\bibfield  {journal} {\bibinfo  {journal} {Applied Physics Letters}\
  }\textbf {\bibinfo {volume} {112}},\ \bibinfo {pages} {042101} (\bibinfo
  {year} {2018})}\BibitemShut {NoStop}%
\bibitem [{\citenamefont {Hollan}\ and\ \citenamefont
  {Durand}(1979)}]{hollan1979fast}%
  \BibitemOpen
  \bibfield  {author} {\bibinfo {author} {\bibfnamefont {L.}~\bibnamefont
  {Hollan}}\ and\ \bibinfo {author} {\bibfnamefont {J.}~\bibnamefont
  {Durand}},\ }\bibfield  {title} {\enquote {\bibinfo {title} {{Fast growth in
  GaAs VPE at low temperature and high partial pressures}},}\ }\href@noop {}
  {\bibfield  {journal} {\bibinfo  {journal} {Journal of Crystal Growth}\
  }\textbf {\bibinfo {volume} {46}},\ \bibinfo {pages} {665--670} (\bibinfo
  {year} {1979})}\BibitemShut {NoStop}%
\bibitem [{\citenamefont {DenBaars}\ \emph {et~al.}(1986)\citenamefont
  {DenBaars}, \citenamefont {Maa}, \citenamefont {Dapkus}, \citenamefont
  {Danner},\ and\ \citenamefont {Lee}}]{denbaars1986homogeneous}%
  \BibitemOpen
  \bibfield  {author} {\bibinfo {author} {\bibfnamefont {S.}~\bibnamefont
  {DenBaars}}, \bibinfo {author} {\bibfnamefont {B.}~\bibnamefont {Maa}},
  \bibinfo {author} {\bibfnamefont {P.}~\bibnamefont {Dapkus}}, \bibinfo
  {author} {\bibfnamefont {A.}~\bibnamefont {Danner}}, \ and\ \bibinfo {author}
  {\bibfnamefont {H.~C.}\ \bibnamefont {Lee}},\ }\bibfield  {title} {\enquote
  {\bibinfo {title} {Homogeneous and heterogeneous thermal decomposition rates
  of trimethylgallium and arsine and their relevance to the growth of gaas by
  mocvd},}\ }\href@noop {} {\bibfield  {journal} {\bibinfo  {journal} {Journal
  of Crystal Growth}\ }\textbf {\bibinfo {volume} {77}},\ \bibinfo {pages}
  {188--193} (\bibinfo {year} {1986})}\BibitemShut {NoStop}%
\bibitem [{\citenamefont {Shaw}(1975)}]{shaw1975kinetic}%
  \BibitemOpen
  \bibfield  {author} {\bibinfo {author} {\bibfnamefont {D.~W.}\ \bibnamefont
  {Shaw}},\ }\bibfield  {title} {\enquote {\bibinfo {title} {{Kinetic aspects
  in the vapour phase epitaxy of III--V compounds}},}\ }\href@noop {}
  {\bibfield  {journal} {\bibinfo  {journal} {Journal of crystal growth}\
  }\textbf {\bibinfo {volume} {31}},\ \bibinfo {pages} {130--141} (\bibinfo
  {year} {1975})}\BibitemShut {NoStop}%
\bibitem [{\citenamefont {Kangawa}\ \emph {et~al.}(2002)\citenamefont
  {Kangawa}, \citenamefont {Ito}, \citenamefont {Hiraoka}, \citenamefont
  {Taguchi}, \citenamefont {Shiraishi},\ and\ \citenamefont
  {Ohachi}}]{kangawa2002theoretical}%
  \BibitemOpen
  \bibfield  {author} {\bibinfo {author} {\bibfnamefont {Y.}~\bibnamefont
  {Kangawa}}, \bibinfo {author} {\bibfnamefont {T.}~\bibnamefont {Ito}},
  \bibinfo {author} {\bibfnamefont {Y.}~\bibnamefont {Hiraoka}}, \bibinfo
  {author} {\bibfnamefont {A.}~\bibnamefont {Taguchi}}, \bibinfo {author}
  {\bibfnamefont {K.}~\bibnamefont {Shiraishi}}, \ and\ \bibinfo {author}
  {\bibfnamefont {T.}~\bibnamefont {Ohachi}},\ }\bibfield  {title} {\enquote
  {\bibinfo {title} {{Theoretical approach to influence of As2 pressure on GaAs
  growth kinetics}},}\ }\href@noop {} {\bibfield  {journal} {\bibinfo
  {journal} {Surface science}\ }\textbf {\bibinfo {volume} {507}},\ \bibinfo
  {pages} {285--289} (\bibinfo {year} {2002})}\BibitemShut {NoStop}%
\bibitem [{\citenamefont {Young}\ \emph {et~al.}(2013)\citenamefont {Young},
  \citenamefont {Ptak}, \citenamefont {Kuech}, \citenamefont {Schulte},\ and\
  \citenamefont {Simon}}]{young2013high}%
  \BibitemOpen
  \bibfield  {author} {\bibinfo {author} {\bibfnamefont {D.~L.}\ \bibnamefont
  {Young}}, \bibinfo {author} {\bibfnamefont {A.~J.}\ \bibnamefont {Ptak}},
  \bibinfo {author} {\bibfnamefont {T.~F.}\ \bibnamefont {Kuech}}, \bibinfo
  {author} {\bibfnamefont {K.}~\bibnamefont {Schulte}}, \ and\ \bibinfo
  {author} {\bibfnamefont {J.~D.}\ \bibnamefont {Simon}},\ }\href@noop {}
  {\enquote {\bibinfo {title} {High throughput semiconductor deposition
  system},}\ } (\bibinfo {year} {2013}),\ \bibinfo {note} {uS Patent App.
  13/895,190}\BibitemShut {NoStop}%
\bibitem [{\citenamefont {Biefeld}, \citenamefont {Koleske},\ and\
  \citenamefont {Cederberg}(2015)}]{biefeld2015science}%
  \BibitemOpen
  \bibfield  {author} {\bibinfo {author} {\bibfnamefont {R.~M.}\ \bibnamefont
  {Biefeld}}, \bibinfo {author} {\bibfnamefont {D.~D.}\ \bibnamefont
  {Koleske}}, \ and\ \bibinfo {author} {\bibfnamefont {J.~G.}\ \bibnamefont
  {Cederberg}},\ }\bibfield  {title} {\enquote {\bibinfo {title} {{The Science
  and Practice of Metal-Organic Vapor Phase Epitaxy (MOVPE)}},}\ }in\
  \href@noop {} {\emph {\bibinfo {booktitle} {Handbook of Crystal Growth}}}\
  (\bibinfo  {publisher} {Elsevier},\ \bibinfo {year} {2015})\ pp.\ \bibinfo
  {pages} {95--160}\BibitemShut {NoStop}%
\bibitem [{\citenamefont {Schulte}\ \emph {et~al.}(2019)\citenamefont
  {Schulte}, \citenamefont {Metaferia}, \citenamefont {Simon},\ and\
  \citenamefont {Ptak}}]{schulte2019uniformity}%
  \BibitemOpen
  \bibfield  {author} {\bibinfo {author} {\bibfnamefont {K.~L.}\ \bibnamefont
  {Schulte}}, \bibinfo {author} {\bibfnamefont {W.}~\bibnamefont {Metaferia}},
  \bibinfo {author} {\bibfnamefont {J.}~\bibnamefont {Simon}}, \ and\ \bibinfo
  {author} {\bibfnamefont {A.~J.}\ \bibnamefont {Ptak}},\ }\bibfield  {title}
  {\enquote {\bibinfo {title} {{Uniformity of \ce{GaAs} solar cells grown in a
  kinetically-limited regime by dynamic hydride vapor phase epitaxy}},}\
  }\href@noop {} {\bibfield  {journal} {\bibinfo  {journal} {Solar Energy
  Materials and Solar Cells}\ }\textbf {\bibinfo {volume} {197}},\ \bibinfo
  {pages} {84--92} (\bibinfo {year} {2019})}\BibitemShut {NoStop}%
\bibitem [{\citenamefont {Schulte}\ \emph {et~al.}(2016)\citenamefont
  {Schulte}, \citenamefont {Simon}, \citenamefont {Jain}, \citenamefont
  {Young},\ and\ \citenamefont {Ptak}}]{schulte2016kinetic}%
  \BibitemOpen
  \bibfield  {author} {\bibinfo {author} {\bibfnamefont {K.~L.}\ \bibnamefont
  {Schulte}}, \bibinfo {author} {\bibfnamefont {J.}~\bibnamefont {Simon}},
  \bibinfo {author} {\bibfnamefont {N.}~\bibnamefont {Jain}}, \bibinfo {author}
  {\bibfnamefont {D.~L.}\ \bibnamefont {Young}}, \ and\ \bibinfo {author}
  {\bibfnamefont {A.~J.}\ \bibnamefont {Ptak}},\ }\bibfield  {title} {\enquote
  {\bibinfo {title} {{A kinetic model for GaAs growth by hydride vapor phase
  epitaxy}},}\ }in\ \href@noop {} {\emph {\bibinfo {booktitle} {2016 IEEE 43rd
  Photovoltaic Specialists Conference (PVSC)}}}\ (\bibinfo {organization}
  {IEEE},\ \bibinfo {year} {2016})\ pp.\ \bibinfo {pages}
  {1930--1933}\BibitemShut {NoStop}%
\bibitem [{\citenamefont {Harrous}\ \emph {et~al.}(1988)\citenamefont
  {Harrous}, \citenamefont {Chaput}, \citenamefont {Bendraoui}, \citenamefont
  {Cadoret}, \citenamefont {Pariset},\ and\ \citenamefont
  {Cadoret}}]{harrous1988phosphine}%
  \BibitemOpen
  \bibfield  {author} {\bibinfo {author} {\bibfnamefont {M.}~\bibnamefont
  {Harrous}}, \bibinfo {author} {\bibfnamefont {L.}~\bibnamefont {Chaput}},
  \bibinfo {author} {\bibfnamefont {A.}~\bibnamefont {Bendraoui}}, \bibinfo
  {author} {\bibfnamefont {M.}~\bibnamefont {Cadoret}}, \bibinfo {author}
  {\bibfnamefont {C.}~\bibnamefont {Pariset}}, \ and\ \bibinfo {author}
  {\bibfnamefont {R.}~\bibnamefont {Cadoret}},\ }\bibfield  {title} {\enquote
  {\bibinfo {title} {{Phosphine and arsine decomposition in CVD reactors for
  InP and InGaAs growth}},}\ }\href@noop {} {\bibfield  {journal} {\bibinfo
  {journal} {Journal of Crystal Growth}\ }\textbf {\bibinfo {volume} {92}},\
  \bibinfo {pages} {423--431} (\bibinfo {year} {1988})}\BibitemShut {NoStop}%
\bibitem [{\citenamefont {Braman}, \citenamefont {Oliver},\ and\ \citenamefont
  {Raman}(2013)}]{braman2013bayesian}%
  \BibitemOpen
  \bibfield  {author} {\bibinfo {author} {\bibfnamefont {K.}~\bibnamefont
  {Braman}}, \bibinfo {author} {\bibfnamefont {T.~A.}\ \bibnamefont {Oliver}},
  \ and\ \bibinfo {author} {\bibfnamefont {V.}~\bibnamefont {Raman}},\
  }\bibfield  {title} {\enquote {\bibinfo {title} {Bayesian analysis of syngas
  chemistry models},}\ }\href@noop {} {\bibfield  {journal} {\bibinfo
  {journal} {Combustion Theory and Modelling}\ }\textbf {\bibinfo {volume}
  {17}},\ \bibinfo {pages} {858--887} (\bibinfo {year} {2013})}\BibitemShut
  {NoStop}%
\bibitem [{\citenamefont {Bell}\ \emph {et~al.}(2019)\citenamefont {Bell},
  \citenamefont {Day}, \citenamefont {Goodman}, \citenamefont {Grout},\ and\
  \citenamefont {Morzfeld}}]{bell2019bayesian}%
  \BibitemOpen
  \bibfield  {author} {\bibinfo {author} {\bibfnamefont {J.}~\bibnamefont
  {Bell}}, \bibinfo {author} {\bibfnamefont {M.}~\bibnamefont {Day}}, \bibinfo
  {author} {\bibfnamefont {J.}~\bibnamefont {Goodman}}, \bibinfo {author}
  {\bibfnamefont {R.}~\bibnamefont {Grout}}, \ and\ \bibinfo {author}
  {\bibfnamefont {M.}~\bibnamefont {Morzfeld}},\ }\bibfield  {title} {\enquote
  {\bibinfo {title} {A bayesian approach to calibrating hydrogen flame kinetics
  using many experiments and parameters},}\ }\href@noop {} {\bibfield
  {journal} {\bibinfo  {journal} {Combustion and Flame}\ }\textbf {\bibinfo
  {volume} {205}},\ \bibinfo {pages} {305--315} (\bibinfo {year}
  {2019})}\BibitemShut {NoStop}%
\bibitem [{\citenamefont {Gr{\"u}ter}\ \emph {et~al.}(1989)\citenamefont
  {Gr{\"u}ter}, \citenamefont {Deschler}, \citenamefont {J{\"u}rgensen},
  \citenamefont {Beccard},\ and\ \citenamefont {Balk}}]{gruter1989deposition}%
  \BibitemOpen
  \bibfield  {author} {\bibinfo {author} {\bibfnamefont {K.}~\bibnamefont
  {Gr{\"u}ter}}, \bibinfo {author} {\bibfnamefont {M.}~\bibnamefont
  {Deschler}}, \bibinfo {author} {\bibfnamefont {H.}~\bibnamefont
  {J{\"u}rgensen}}, \bibinfo {author} {\bibfnamefont {R.}~\bibnamefont
  {Beccard}}, \ and\ \bibinfo {author} {\bibfnamefont {P.}~\bibnamefont
  {Balk}},\ }\bibfield  {title} {\enquote {\bibinfo {title} {{Deposition of
  high quality GaAs films at fast rates in the LP-CVD system}},}\ }\href@noop
  {} {\bibfield  {journal} {\bibinfo  {journal} {Journal of crystal growth}\
  }\textbf {\bibinfo {volume} {94}},\ \bibinfo {pages} {607--612} (\bibinfo
  {year} {1989})}\BibitemShut {NoStop}%
\bibitem [{\citenamefont {Pedregosa}\ \emph {et~al.}(2011)\citenamefont
  {Pedregosa}, \citenamefont {Varoquaux}, \citenamefont {Gramfort},
  \citenamefont {Michel}, \citenamefont {Thirion}, \citenamefont {Grisel},
  \citenamefont {Blondel}, \citenamefont {Prettenhofer}, \citenamefont {Weiss},
  \citenamefont {Dubourg}, \citenamefont {Vanderplas}, \citenamefont {Passos},
  \citenamefont {Cournapeau}, \citenamefont {Brucher}, \citenamefont {Perrot},\
  and\ \citenamefont {Duchesnay}}]{scikit-learn}%
  \BibitemOpen
  \bibfield  {author} {\bibinfo {author} {\bibfnamefont {F.}~\bibnamefont
  {Pedregosa}}, \bibinfo {author} {\bibfnamefont {G.}~\bibnamefont
  {Varoquaux}}, \bibinfo {author} {\bibfnamefont {A.}~\bibnamefont {Gramfort}},
  \bibinfo {author} {\bibfnamefont {V.}~\bibnamefont {Michel}}, \bibinfo
  {author} {\bibfnamefont {B.}~\bibnamefont {Thirion}}, \bibinfo {author}
  {\bibfnamefont {O.}~\bibnamefont {Grisel}}, \bibinfo {author} {\bibfnamefont
  {M.}~\bibnamefont {Blondel}}, \bibinfo {author} {\bibfnamefont
  {P.}~\bibnamefont {Prettenhofer}}, \bibinfo {author} {\bibfnamefont
  {R.}~\bibnamefont {Weiss}}, \bibinfo {author} {\bibfnamefont
  {V.}~\bibnamefont {Dubourg}}, \bibinfo {author} {\bibfnamefont
  {J.}~\bibnamefont {Vanderplas}}, \bibinfo {author} {\bibfnamefont
  {A.}~\bibnamefont {Passos}}, \bibinfo {author} {\bibfnamefont
  {D.}~\bibnamefont {Cournapeau}}, \bibinfo {author} {\bibfnamefont
  {M.}~\bibnamefont {Brucher}}, \bibinfo {author} {\bibfnamefont
  {M.}~\bibnamefont {Perrot}}, \ and\ \bibinfo {author} {\bibfnamefont
  {E.}~\bibnamefont {Duchesnay}},\ }\bibfield  {title} {\enquote {\bibinfo
  {title} {{Scikit-learn: Machine Learning in {P}ython}},}\ }\href@noop {}
  {\bibfield  {journal} {\bibinfo  {journal} {Journal of Machine Learning
  Research}\ }\textbf {\bibinfo {volume} {12}},\ \bibinfo {pages} {2825--2830}
  (\bibinfo {year} {2011})}\BibitemShut {NoStop}%
\bibitem [{\citenamefont {Hassanaly}\ \emph {et~al.}(2018)\citenamefont
  {Hassanaly}, \citenamefont {Koo}, \citenamefont {Lietz}, \citenamefont
  {Chong},\ and\ \citenamefont {Raman}}]{hassanaly2018minimally}%
  \BibitemOpen
  \bibfield  {author} {\bibinfo {author} {\bibfnamefont {M.}~\bibnamefont
  {Hassanaly}}, \bibinfo {author} {\bibfnamefont {H.}~\bibnamefont {Koo}},
  \bibinfo {author} {\bibfnamefont {C.~F.}\ \bibnamefont {Lietz}}, \bibinfo
  {author} {\bibfnamefont {S.~T.}\ \bibnamefont {Chong}}, \ and\ \bibinfo
  {author} {\bibfnamefont {V.}~\bibnamefont {Raman}},\ }\bibfield  {title}
  {\enquote {\bibinfo {title} {{A minimally-dissipative low-Mach number solver
  for complex reacting flows in OpenFOAM}},}\ }\href@noop {} {\bibfield
  {journal} {\bibinfo  {journal} {Computers \& Fluids}\ }\textbf {\bibinfo
  {volume} {162}},\ \bibinfo {pages} {11--25} (\bibinfo {year}
  {2018})}\BibitemShut {NoStop}%
\bibitem [{\citenamefont {Hassanaly}\ \emph {et~al.}(2020)\citenamefont
  {Hassanaly}, \citenamefont {Tang}, \citenamefont {Barwey},\ and\
  \citenamefont {Raman}}]{hassanaly2020data}%
  \BibitemOpen
  \bibfield  {author} {\bibinfo {author} {\bibfnamefont {M.}~\bibnamefont
  {Hassanaly}}, \bibinfo {author} {\bibfnamefont {Y.}~\bibnamefont {Tang}},
  \bibinfo {author} {\bibfnamefont {S.}~\bibnamefont {Barwey}}, \ and\ \bibinfo
  {author} {\bibfnamefont {V.}~\bibnamefont {Raman}},\ }\bibfield  {title}
  {\enquote {\bibinfo {title} {{Data-driven Analysis of Relight variability of
  Jet Fuels induced by Turbulence}},}\ }\href@noop {} {\bibfield  {journal}
  {\bibinfo  {journal} {Combustion and Flame}\ }\textbf {\bibinfo {volume}
  {225}},\ \bibinfo {pages} {453--467} (\bibinfo {year} {2020})}\BibitemShut
  {NoStop}%
\bibitem [{\citenamefont {Tang}\ and\ \citenamefont
  {Raman}(2021)}]{tang2021large}%
  \BibitemOpen
  \bibfield  {author} {\bibinfo {author} {\bibfnamefont {Y.}~\bibnamefont
  {Tang}}\ and\ \bibinfo {author} {\bibfnamefont {V.}~\bibnamefont {Raman}},\
  }\bibfield  {title} {\enquote {\bibinfo {title} {Large eddy simulation of
  premixed turbulent combustion using a non-adiabatic, strain-sensitive
  flamelet approach},}\ }\href@noop {} {\bibfield  {journal} {\bibinfo
  {journal} {Combustion and Flame}\ }\textbf {\bibinfo {volume} {234}},\
  \bibinfo {pages} {111655} (\bibinfo {year} {2021})}\BibitemShut {NoStop}%
\bibitem [{\citenamefont {Koo}\ \emph {et~al.}(2017)\citenamefont {Koo},
  \citenamefont {Hassanaly}, \citenamefont {Raman}, \citenamefont {Mueller},\
  and\ \citenamefont {Peter~Geigle}}]{koo2017large}%
  \BibitemOpen
  \bibfield  {author} {\bibinfo {author} {\bibfnamefont {H.}~\bibnamefont
  {Koo}}, \bibinfo {author} {\bibfnamefont {M.}~\bibnamefont {Hassanaly}},
  \bibinfo {author} {\bibfnamefont {V.}~\bibnamefont {Raman}}, \bibinfo
  {author} {\bibfnamefont {M.~E.}\ \bibnamefont {Mueller}}, \ and\ \bibinfo
  {author} {\bibfnamefont {K.}~\bibnamefont {Peter~Geigle}},\ }\bibfield
  {title} {\enquote {\bibinfo {title} {Large-eddy simulation of soot formation
  in a model gas turbine combustor},}\ }\href@noop {} {\bibfield  {journal}
  {\bibinfo  {journal} {Journal of Engineering for Gas Turbines and Power}\
  }\textbf {\bibinfo {volume} {139}} (\bibinfo {year} {2017})}\BibitemShut
  {NoStop}%
\bibitem [{\citenamefont {Sitaraman}\ \emph {et~al.}(2021)\citenamefont
  {Sitaraman}, \citenamefont {Yellapantula}, \citenamefont {de~Frahan},
  \citenamefont {Perry}, \citenamefont {Rood}, \citenamefont {Grout},\ and\
  \citenamefont {Day}}]{sitaraman2021adaptive}%
  \BibitemOpen
  \bibfield  {author} {\bibinfo {author} {\bibfnamefont {H.}~\bibnamefont
  {Sitaraman}}, \bibinfo {author} {\bibfnamefont {S.}~\bibnamefont
  {Yellapantula}}, \bibinfo {author} {\bibfnamefont {M.~T.~H.}\ \bibnamefont
  {de~Frahan}}, \bibinfo {author} {\bibfnamefont {B.}~\bibnamefont {Perry}},
  \bibinfo {author} {\bibfnamefont {J.}~\bibnamefont {Rood}}, \bibinfo {author}
  {\bibfnamefont {R.}~\bibnamefont {Grout}}, \ and\ \bibinfo {author}
  {\bibfnamefont {M.}~\bibnamefont {Day}},\ }\bibfield  {title} {\enquote
  {\bibinfo {title} {Adaptive mesh based combustion simulations of direct fuel
  injection effects in a supersonic cavity flame-holder},}\ }\href@noop {}
  {\bibfield  {journal} {\bibinfo  {journal} {Combustion and Flame}\ }\textbf
  {\bibinfo {volume} {232}},\ \bibinfo {pages} {111531} (\bibinfo {year}
  {2021})}\BibitemShut {NoStop}%
\bibitem [{\citenamefont {Smagorinsky}(1963)}]{smagorinsky1963general}%
  \BibitemOpen
  \bibfield  {author} {\bibinfo {author} {\bibfnamefont {J.}~\bibnamefont
  {Smagorinsky}},\ }\bibfield  {title} {\enquote {\bibinfo {title} {{General
  circulation experiments with the primitive equations: I. The basic
  experiment}},}\ }\href@noop {} {\bibfield  {journal} {\bibinfo  {journal}
  {Monthly weather review}\ }\textbf {\bibinfo {volume} {91}},\ \bibinfo
  {pages} {99--164} (\bibinfo {year} {1963})}\BibitemShut {NoStop}%
\bibitem [{\citenamefont {Tang}\ \emph {et~al.}(2019)\citenamefont {Tang},
  \citenamefont {Hassanaly}, \citenamefont {Raman}, \citenamefont {Sforzo},\
  and\ \citenamefont {Seitzman}}]{tang2019comprehensive}%
  \BibitemOpen
  \bibfield  {author} {\bibinfo {author} {\bibfnamefont {Y.}~\bibnamefont
  {Tang}}, \bibinfo {author} {\bibfnamefont {M.}~\bibnamefont {Hassanaly}},
  \bibinfo {author} {\bibfnamefont {V.}~\bibnamefont {Raman}}, \bibinfo
  {author} {\bibfnamefont {B.}~\bibnamefont {Sforzo}}, \ and\ \bibinfo {author}
  {\bibfnamefont {J.}~\bibnamefont {Seitzman}},\ }\bibfield  {title} {\enquote
  {\bibinfo {title} {A comprehensive modeling procedure for estimating
  statistical properties of forced ignition},}\ }\href@noop {} {\bibfield
  {journal} {\bibinfo  {journal} {Combustion and Flame}\ }\textbf {\bibinfo
  {volume} {206}},\ \bibinfo {pages} {158--176} (\bibinfo {year}
  {2019})}\BibitemShut {NoStop}%
\bibitem [{\citenamefont {Maestri}\ and\ \citenamefont
  {Cuoci}(2013)}]{maestri2013coupling}%
  \BibitemOpen
  \bibfield  {author} {\bibinfo {author} {\bibfnamefont {M.}~\bibnamefont
  {Maestri}}\ and\ \bibinfo {author} {\bibfnamefont {A.}~\bibnamefont
  {Cuoci}},\ }\bibfield  {title} {\enquote {\bibinfo {title} {{Coupling CFD
  with detailed microkinetic modeling in heterogeneous catalysis}},}\
  }\href@noop {} {\bibfield  {journal} {\bibinfo  {journal} {Chemical
  Engineering Science}\ }\textbf {\bibinfo {volume} {96}},\ \bibinfo {pages}
  {106--117} (\bibinfo {year} {2013})}\BibitemShut {NoStop}%
\bibitem [{\citenamefont {Strang}(1968)}]{strang1968construction}%
  \BibitemOpen
  \bibfield  {author} {\bibinfo {author} {\bibfnamefont {G.}~\bibnamefont
  {Strang}},\ }\bibfield  {title} {\enquote {\bibinfo {title} {On the
  construction and comparison of difference schemes},}\ }\href@noop {}
  {\bibfield  {journal} {\bibinfo  {journal} {SIAM journal on numerical
  analysis}\ }\textbf {\bibinfo {volume} {5}},\ \bibinfo {pages} {506--517}
  (\bibinfo {year} {1968})}\BibitemShut {NoStop}%
\bibitem [{\citenamefont {Heye}, \citenamefont {Raman},\ and\ \citenamefont
  {Masri}(2013)}]{heye2013probability}%
  \BibitemOpen
  \bibfield  {author} {\bibinfo {author} {\bibfnamefont {C.}~\bibnamefont
  {Heye}}, \bibinfo {author} {\bibfnamefont {V.}~\bibnamefont {Raman}}, \ and\
  \bibinfo {author} {\bibfnamefont {A.~R.}\ \bibnamefont {Masri}},\ }\bibfield
  {title} {\enquote {\bibinfo {title} {{LES/probability density function
  approach for the simulation of an ethanol spray flame}},}\ }\href@noop {}
  {\bibfield  {journal} {\bibinfo  {journal} {Proceedings of the Combustion
  Institute}\ }\textbf {\bibinfo {volume} {34}},\ \bibinfo {pages} {1633--1641}
  (\bibinfo {year} {2013})}\BibitemShut {NoStop}%
\bibitem [{\citenamefont {Chong}\ \emph {et~al.}(2018)\citenamefont {Chong},
  \citenamefont {Hassanaly}, \citenamefont {Koo}, \citenamefont {Mueller},
  \citenamefont {Raman},\ and\ \citenamefont {Geigle}}]{chong2018large}%
  \BibitemOpen
  \bibfield  {author} {\bibinfo {author} {\bibfnamefont {S.~T.}\ \bibnamefont
  {Chong}}, \bibinfo {author} {\bibfnamefont {M.}~\bibnamefont {Hassanaly}},
  \bibinfo {author} {\bibfnamefont {H.}~\bibnamefont {Koo}}, \bibinfo {author}
  {\bibfnamefont {M.~E.}\ \bibnamefont {Mueller}}, \bibinfo {author}
  {\bibfnamefont {V.}~\bibnamefont {Raman}}, \ and\ \bibinfo {author}
  {\bibfnamefont {K.-P.}\ \bibnamefont {Geigle}},\ }\bibfield  {title}
  {\enquote {\bibinfo {title} {Large eddy simulation of pressure and
  dilution-jet effects on soot formation in a model aircraft swirl
  combustor},}\ }\href@noop {} {\bibfield  {journal} {\bibinfo  {journal}
  {Combustion and Flame}\ }\textbf {\bibinfo {volume} {192}},\ \bibinfo {pages}
  {452--472} (\bibinfo {year} {2018})}\BibitemShut {NoStop}%
\bibitem [{\citenamefont {Oehlschlaeger}\ \emph {et~al.}(2009)\citenamefont
  {Oehlschlaeger}, \citenamefont {Steinberg}, \citenamefont {Westbrook},\ and\
  \citenamefont {Pitz}}]{oehlschlaeger2009autoignition}%
  \BibitemOpen
  \bibfield  {author} {\bibinfo {author} {\bibfnamefont {M.~A.}\ \bibnamefont
  {Oehlschlaeger}}, \bibinfo {author} {\bibfnamefont {J.}~\bibnamefont
  {Steinberg}}, \bibinfo {author} {\bibfnamefont {C.~K.}\ \bibnamefont
  {Westbrook}}, \ and\ \bibinfo {author} {\bibfnamefont {W.~J.}\ \bibnamefont
  {Pitz}},\ }\bibfield  {title} {\enquote {\bibinfo {title} {{The autoignition
  of iso-cetane at high to moderate temperatures and elevated pressures: Shock
  tube experiments and kinetic modeling}},}\ }\href@noop {} {\bibfield
  {journal} {\bibinfo  {journal} {Combustion and flame}\ }\textbf {\bibinfo
  {volume} {156}},\ \bibinfo {pages} {2165--2172} (\bibinfo {year}
  {2009})}\BibitemShut {NoStop}%
\bibitem [{\citenamefont {Kee}\ \emph {et~al.}(1999)\citenamefont {Kee},
  \citenamefont {Dixon-Lewis}, \citenamefont {Warnatz}, \citenamefont
  {Coltrin}, \citenamefont {Miller},\ and\ \citenamefont
  {Moffat}}]{kee1999transport}%
  \BibitemOpen
  \bibfield  {author} {\bibinfo {author} {\bibfnamefont {R.}~\bibnamefont
  {Kee}}, \bibinfo {author} {\bibfnamefont {G.}~\bibnamefont {Dixon-Lewis}},
  \bibinfo {author} {\bibfnamefont {J.}~\bibnamefont {Warnatz}}, \bibinfo
  {author} {\bibfnamefont {M.}~\bibnamefont {Coltrin}}, \bibinfo {author}
  {\bibfnamefont {J.}~\bibnamefont {Miller}}, \ and\ \bibinfo {author}
  {\bibfnamefont {H.}~\bibnamefont {Moffat}},\ }\bibfield  {title} {\enquote
  {\bibinfo {title} {Transport: a software package for the evaluation of
  gas-phase, multicomponent transport properties},}\ }\href@noop {} {\bibfield
  {journal} {\bibinfo  {journal} {Chemkin Collection}\ } (\bibinfo {year}
  {1999})}\BibitemShut {NoStop}%
\bibitem [{\citenamefont {Bergman}\ \emph {et~al.}(2011)\citenamefont
  {Bergman}, \citenamefont {Incropera}, \citenamefont {DeWitt},\ and\
  \citenamefont {Lavine}}]{bergman2011fundamentals}%
  \BibitemOpen
  \bibfield  {author} {\bibinfo {author} {\bibfnamefont {T.~L.}\ \bibnamefont
  {Bergman}}, \bibinfo {author} {\bibfnamefont {F.~P.}\ \bibnamefont
  {Incropera}}, \bibinfo {author} {\bibfnamefont {D.~P.}\ \bibnamefont
  {DeWitt}}, \ and\ \bibinfo {author} {\bibfnamefont {A.~S.}\ \bibnamefont
  {Lavine}},\ }\href@noop {} {\emph {\bibinfo {title} {Fundamentals of heat and
  mass transfer}}}\ (\bibinfo  {publisher} {John Wiley \& Sons},\ \bibinfo
  {year} {2011})\BibitemShut {NoStop}%
\end{thebibliography}

\end{document}